\def\upd{{\rm d}}

\documentclass[pre,twocolumn,showpacs]{revtex4}
\usepackage{graphicx}
\usepackage{amsbsy}

\begin{document}

\title{Steady state work fluctuations of a dragged particle under external and thermal noise}

\author{A. Baule and E. G. D. Cohen}

\affiliation{                    
The Rockefeller University, 1230 York Avenue, New York, NY 10065, USA
}

\begin{abstract}

We consider a particle, confined to a moving harmonic potential, under the influence of friction and external asymmetric Poissonian shot noise (PSN). We study the fluctuations of the work done to maintain this system in a nonequilibrium steady state. PSN generalizes the usual Gaussian noise and can be considered to be a paradigm of external noise, where fluctuation and dissipation originate from physically independent mechanisms. We consider two scenarios: (i) the noise is given purely by PSN and (ii) in addition to PSN the particle is subject to white Gaussian noise. In both cases we derive exact expressions for the large deviation form of the work distribution, which are characterized by the time scales of the system. We show that the usual steady state fluctuation theorem is violated in our model and that in a certain parameter regime large negative work fluctuations are more likely to occur than the corresponding positive ones, though the average work is always positive.

\end{abstract}

\date{\today}


\maketitle

\section{Introduction}

An understanding of the fluctuation properties of systems away from thermal equilibrium plays an important role for a statistical mechanical theory of these systems. Nonequilibrium fluctuations can have significant effects in small scale systems from physics to biology, which become more and more important in technological innovations. However, from a fundamental theoretical point of view, not much is known about the general aspects of nonequilibrium fluctuations, contrary to our understanding of equilibrium fluctuations.

The simplest generalization of the equilibrium state is a nonequilibrium steady state (NESS), which arises physically due to a balance between driving forces and dissipative forces acting on the system. One of the simplest nonequilibrium systems that can be maintained in a NESS is a Brownian particle, e.g., a spherical colloidal particle in a fluid, confined to a harmonic potential which moves with constant velocity. In addition to the potential force the particle is subject to friction and thermal noise from the surrounding fluid. In this system the driving is due to the time-dependent force from the moving potential and the dissipation is due to the friction of the particle in the fluid. In order to maintain the NESS one has to perform work on the particle, which, in turn, is partly dissipated in the fluid as heat, which has to be removed, and partly stored as potential energy in the harmonic potential. Macroscopically, both the work done on the system as well as the heat removed from it have to be positive in the NESS, while deviations from this macroscopic behavior arise due to fluctuations.

The properties of the work fluctuations in this dragged particle model have been investigated experimentally and theoretically in the literature (see e.g. \cite{Wang02,VanZon03,Taniguchi07,Taniguchi08}), in particular with respect to the validity of the so-called steady state fluctuation theorem (SSFT) (cf. \cite{EvansD93,Gallavotti95,Gallavotti95b,Kurchan98,Lebowitz99}). This theorem states that the probability distribution $\Pi_\tau(p)$ of observing a particular value of a scaled dimensionless work value $p$ over time $\tau$ satisfies a certain symmetry relation, which can be formulated as \cite{Footnote1}
\begin{eqnarray}
\label{conventional}
\frac{\Pi_\tau(p)}{\Pi_\tau(-p)}\cong e^{c\tau p},
\end{eqnarray}
where $\cong$ indicates the asymptotic behavior for large $\tau$ and $c$ is a constant. Eq.~(\ref{conventional}) represents a refinement of the second law in that it quantifies the probability of observing temporary second law violations (negative $p$) in the NESS. The SSFT Eq.~(\ref{conventional}) for the work fluctuations has indeed been validated in the dragged particle model if the noise is modeled as thermal Gaussian noise \cite{Wang02,VanZon03}. Yet, if the noise is modeled as L\'evy noise the work fluctuations violate the SSFT and one can show that an \textit{anomalous fluctuation theorem} \cite{Touchette07,Touchette09} holds. This result highlights the importance of explicitly verifying a general result, such as Eq.~(\ref{conventional}), in concrete models.

In the present article we investigate the work fluctuation properties of the above mentioned NESS model when the noise from the environment has the characteristics of external asymmetric Poissonian shot noise (PSN), i.e., it is given as a sequence of one-sided Poisson distributed pulses with random amplitudes \cite{Feynman}. PSN is a natural description of fluctuations in nature \cite{Feynman,VanKampen} and is ubiquitous, e.g., in physics, electric engineering, and biology. Importantly, PSN generalizes the usual Gaussian noise, to which it converges in a certain limit, and is a paradigm of \textit{external noise}: the dissipation and the noise originate from physically independent mechanisms \cite{VanKampen}. This implies that the noise strength of PSN does not have to be related to the friction by a fluctuation-dissipation relation, contrary to thermal noise in the case of Brownian motion. 

PSN allows us to investigate a number of important physical concepts in the context of nonequilibrium fluctuations: (i) \textit{Time scales}: PSN introduces additional characteristic times in the system which crucially influence its fluctuation behavior; (ii) \textit{Symmetry} of the noise: PSN is usually asymmetric (one-sided); (iii) \textit{Singularities}: the asymmetry of the noise gives rise to an effective interaction between the noise and the potential, leading to an effectively nonlinear system with singular features. Nevertheless, the work distribution can be calculated in analytical form in our model and is completely characterized by its time scales. The interplay of these time scales leads to certain transitions in the behavior of the work fluctuations, as discussed in a recent Rapid Communication \cite{Baule09}. Here, we present the full theory of Ref.~\cite{Baule09} including inertial effects. Moreover, we consider the case where thermal Gaussian noise is superimposed on the PSN. This Gaussian noise models the effect of additional thermal noise on the particle, which might be relevant in an experimental realization of our model.

The article is organized as follows. In the next section \ref{Sec_model} we introduce the nonequilibrium particle model under the influence of one-sided PSN. Symmetries, time scales, and singularities of this model are discussed in Sec.~\ref{Sec_sym}. The characteristic function of the work fluctuations including inertia is calculated in Sec.~\ref{Sec_cf}. In the overdamped regime the characteristic function simplifies significantly and an analytical treatment is possible. This allows us to analytically determine the distribution of the work fluctuations using the method of steepest descent and investigate the fluctuation properties in Sec.~\ref{Sec_FT}. If the effect of inertia is non-negligible the work fluctuations exhibit time-oscillatory behavior, as studied in Sec.~\ref{Sec_inertia}. Furthermore, in Sec.~\ref{Sec_GP} we discuss the combined effect of PSN and thermal Gaussian noise on the work fluctuations. We close with some concluding remarks in Sec.~\ref{Sec_C}.

\section{A nonequilibrium stochastic particle model with external PSN}
\label{Sec_model}

We consider a particle which moves in a time-dependent harmonic potential under the influence of friction and \textit{external noise}. The basic equation of motion for the position $x(t)$ of the particle of mass $m$ in the laboratory frame reads (cf. \cite{Taniguchi07})
\begin{eqnarray}
\label{trap}
m\ddot{x}(t)+\alpha \dot{x}(t)=-\kappa(x(t)-vt)+\xi(t).
\end{eqnarray}
Here, the force $-\kappa(x(t)-vt)$ stems from a particle confining potential $U(x,t)=\kappa(x-vt)^2/2$ which is pulled with constant velocity $v$. For our purposes it is sufficient to restrict the discussion to one dimension. The parameter $\kappa$ denotes the strength of the potential, $\alpha$ the friction coefficient, and $\xi(t)$ stochastic noise due to the environment, to be defined below (cf. Eq.~(\ref{xi})). If the potential moves for a time period $\tau$, a certain amount of work is done on the particle, namely
\begin{eqnarray}
\label{work}
W_\tau=-\kappa v\int_0^\tau (x(t)-vt) \upd t.
\end{eqnarray}
In an experimental setup, where the harmonic potential can be induced e.g. via lasers \cite{Wang02}, Eq.~(\ref{work}) is the mechanical work needed in order to move the potential in time $\tau$. If the potential is stationary (i.e. $v=0$), no work is performed at all. This means that we ignore the work originating from the stochastic motion of the particle, which is attributed to the heat \cite{VanZon03,Sekimoto98}.

The work is partly dissipated as heat due to particle friction, and partly stored as potential energy in the potential. In the steady state the mean value of the work $\left<W_\tau\right>$ is positive, since we have to perform a macroscopic amount of work against the friction in order to maintain a NESS. This is basically a statement of the second law of thermodynamics (cf. \cite{Taniguchi08b,Cohen08}).

The mathematical treatment of our model Eq.~(\ref{trap}) is simplified if we transform to a coordinate system in a comoving frame. Let us denote the position of the particle in the comoving frame by $y(t)\equiv x(t)-vt$. The equation of motion for $y(t)$ then reads
\begin{eqnarray}
\label{oscillator2}
m\ddot{y}(t)+\alpha\dot{y}(t)=-\kappa y(t)-\alpha v+\xi(t),
\end{eqnarray}
and the work is given by
\begin{eqnarray}
\label{work_par}
W_\tau=-\kappa v\int_0^\tau y(t)\upd t.
\end{eqnarray}
Although the work is expressed in terms of the comoving coordinate $y(t)$, Eq.~(\ref{work_par}) actually gives the work done in the laboratory frame \cite{Taniguchi07,Taniguchi08}, in whose fluctuations we are interested here.

We consider the model Eq.~(\ref{trap}) under the influence of external PSN rather than the usual Gaussian noise, so that the motion of the particle differs in general from Brownian motion. PSN is specified by a sequence of delta shaped pulses with random amplitudes $\Gamma_k$ and can be expressed in the form \cite{Feynman}
\begin{eqnarray}
\label{F_poiss_noise}
z(t)=\sum_{k=1}^{n_t}\Gamma_k\delta(t-t_k),
\end{eqnarray}  
where $n_t$, the number of delta shaped pulses in time $t$, is determined by the Poisson counting process
\begin{eqnarray}
P(n_t)=\frac{(\lambda t)^n}{n!} e^{-\lambda t}.
\end{eqnarray}
The parameter $\lambda$ denotes the mean number of pulses per unit time (rate of pulses) so that there are $\lambda t$ pulses occurring in the time interval $[0,t]$ on average. In that interval, the time of the $k$th pulse is uniformly distributed. When a pulse occurs, its amplitude $\Gamma_k$ is sampled randomly from a distribution $\rho(\Gamma)$. For $\rho(\Gamma)$ we choose an exponential distribution
\begin{eqnarray}
\label{rho}
\rho(\Gamma)=\frac{1}{\Gamma_0}e^{-\Gamma/\Gamma_0},
\end{eqnarray}
where all amplitudes $\Gamma$ are assumed here to be positive, i.e., the PSN that we consider is \textit{one-sided}.

The noise $z(t)$ specified according to Eq.~(\ref{F_poiss_noise}) has the characteristic functional \cite{Feynman}
\begin{eqnarray}
G_{z(t)}[g(t)]&=&e^{\lambda\int_0^\infty\left(\int_0^\infty e^{i\Gamma g(t)}\rho(\Gamma)\upd\Gamma-1\right)\upd t},
\end{eqnarray}
for a general test function $g(t)$. If $\rho(\Gamma)$ is given by the exponential distribution Eq.~(\ref{rho}) one obtains 
\begin{eqnarray}
\label{noise_functional}
G_{z(t)}[g(t)]&=&e^{\lambda\int_0^\infty\left(\frac{1}{1-i\Gamma_0g(t)}-1\right)\upd t}.
\end{eqnarray}
This characteristic functional implies delta-correlated cumulants \cite{VanKampen80}, which arise due to the delta shape of the stochastic pulses in Eq.~(\ref{F_poiss_noise}):
\begin{eqnarray}
c_n(t_1,...,t_n)&\equiv&\left.\frac{1}{i^n}\frac{\delta^n}{\delta g(t_1)\cdots \delta g(t_n)}\ln G_{z(t)}[g(t)]\right|_{g(t)=0}\nonumber\\
&=&n!\lambda\Gamma_0^n \delta(t_1-t_2)\cdots \delta(t_{n-1}-t_n).
\end{eqnarray}
The first two cumulants, the mean and the variance of $z(t)$, are therefore given by
\begin{eqnarray}
\label{PSN_noise1}
\left<z(t)\right>&=&\lambda\Gamma_0,\\ \label{PSN_noise2}
\left<z(t_1)z(t_2)\right>-\left<z(t_1)\right>\left<z(t_2)\right>&=&2\lambda\Gamma_0^2 \delta(t_1-t_2).
\end{eqnarray}
Here, the brackets $\left<...\right>$ denote the usual ensemble average, which, more precisely, represents a path-integral average with respect to the probability $\mathcal{P}[z(t)]$ of a noise trajectory $z(t)$ (cf. \cite{Taniguchi07}). Since it is convenient to have a noise with zero mean, we take in Eq.~(\ref{trap}) for $\xi(t)$
\begin{eqnarray}
\label{xi}
\xi(t)\equiv z(t)-\lambda\Gamma_0.
\end{eqnarray}
This means that the noise in our model is considered to consist of a random shot noise part and a deterministic part that is equivalent to a constant negative drift force on the particle. Since the shot noise acts only one-sided by our convention, the noise $\xi(t)$ is strongly asymmetric, even though its mean value is zero by construction.

In contrast to the case of a Brownian particle, the noise is here \textit{external}, which implies that the friction coefficient $\alpha$ is not related to the noise strength by a fluctuation-dissipation relation. In the absence of driving (i.e. $v=0$) the stationary distribution of the particle position is therefore not a thermal equilibrium distribution in general. However, in a certain limit the $\xi(t)$ of Eq.~(\ref{xi}) does become Gaussian noise, namely if we take the limits
\begin{eqnarray}
\label{G_limit}
\lambda\rightarrow\infty,\qquad
\Gamma_0\rightarrow 0,
\end{eqnarray}
while keeping
\begin{eqnarray}
\lambda\Gamma_0^2=const.
\end{eqnarray}
If this Gaussian noise is considered to be thermal, i.e., originating from an equilibrium heat bath, the fluctuation-dissipation theorem requires that this constant is given by
\begin{eqnarray}
\label{G_limit_th}
\lambda\Gamma_0^2=\alpha\beta^{-1},
\end{eqnarray}
where $\beta$ can be interpreted as the inverse temperature of the heat bath. This leads to
\begin{eqnarray}
\left<\xi(t)\right>&=&0,\\
\left<\xi(t)\xi(t')\right>&=&2\alpha\beta^{-1}\delta(t-t'),
\end{eqnarray}
while all higher-order cumulants of $\xi(t)$ are zero. In the limits of Eqs.~(\ref{G_limit})---(\ref{G_limit_th}), $\xi(t)$ then represents the standard Gaussian white noise of an equilibrium heat bath.

The limit Eq.~(\ref{G_limit}) means that both the waiting time between successive pulses ($\lambda^{-1}$) and their amplitudes ($\Gamma_k$) become very small, i.e., we have very frequent (at every `time step'), independent, and very small pulses. This is the signature of Gaussian noise. The system can then indeed reach a state of thermal equilibrium if $v=0$.

\section{Symmetries, time scales, and singularities}
\label{Sec_sym}

The system given by Eq.~(\ref{trap}) describes a damped harmonic oscillator with inertia, driven by a time-dependent force and by noise. For symmetric noise (e.g. Gaussian noise) the properties of the work fluctuations in this system are symmetric with respect to the direction of the pulling velocity $v$, i.e., the work distribution can only depend on the absolute value $|v|$. However, the presence of asymmetric shot noise, prescribed by Eqs.~(\ref{F_poiss_noise}) and (\ref{rho}), breaks this symmetry. Since both the direction of $v$ and the direction of the noise (given by the sign of $\Gamma_0$) can in principle be either positive or negative, there are in total four different combinations of the two. It is easy to see that the case $v>0$ and $\Gamma_0>0$ is symmetric with the case $v<0$ and $\Gamma_0<0$. Likewise, the case $v>0$ and $\Gamma_0<0$ is symmetric with the case $v<0$ and $\Gamma_0>0$. Therefore it is sufficient to discuss only two of the four different cases. In the following we investigate the work fluctuations for $v>0$ and $v<0$, while $\Gamma_0>0$ always. 

The oscillator itself is completely characterized by the two time scales
\begin{eqnarray}
\label{tau_m}
\tau_m\equiv m/\alpha,
\end{eqnarray}
the inertial time, and
\begin{eqnarray}
\label{tau_r}
\tau_r\equiv\alpha/\kappa,
\end{eqnarray}
the relaxation time. Associated with the PSN are two additional characteristic time scales. Firstly, we have the time $\tau_\lambda$ defined as
\begin{eqnarray}
\label{tau_l}
\tau_\lambda\equiv\lambda^{-1},
\end{eqnarray}
which is the mean waiting time between two successive pulses. Secondly, we can identify a time scale $\tau_p$, defined as
\begin{eqnarray}
\label{tau_p}
\tau_p\equiv\frac{\Gamma_0}{\alpha|v|},
\end{eqnarray}
relating the mean amplitude of the pulses and the friction due to the driving, so that $\tau_p$ is the ratio of two independent external forces, due to the noise ($\Gamma_0$) and due to the driving ($v$), respectively. In total there are therefore four different time scales in our model: the inertial time $\tau_m$, the relaxation time $\tau_r$, the mean waiting time $\tau_\lambda$ and $\tau_p$. While the first two are intrinsic time scales of the oscillator, the latter two arise due to the particular type of external noise that we consider. We will see below that the four time scales $\tau_m$, $\tau_r$, $\tau_\lambda$, and $\tau_p$ fully specify the properties of the work fluctuations in our model, and that the interplay of these times induces transitions in the qualitative behavior of the work fluctuations.

One familiar example of such a transitional behavior is due to the interplay of the inertial and the relaxation times, Eqs.~(\ref{tau_m}) and (\ref{tau_r}), respectively. Using $\tau_m$ and $\tau_r$ we can express the eigenvalues of the oscillator Eq.~(\ref{trap}) in the form
\begin{eqnarray}
\label{eigenvalues}
\nu_{1,2}=\frac{1}{2\tau_m}\left(-1\pm\sqrt{1-4\tau_m/\tau_r}\right),
\end{eqnarray}
where the index $1$ corresponds to the $+$ sign and $2$ to the $-$ sign, respectively.
We can thus formulate a critical condition (cf. \cite{Taniguchi08})
\begin{eqnarray}
\tau_m=\tau_r/4
\end{eqnarray}
for the transition from real to complex eigenvalues. If $\tau_m>\tau_r/4$ the eigenfrequencies are complex and the influence of the inertia on the dynamics manifests itself in time oscillatory behavior. Moreover, in a regime where $\tau_m\ll \tau_r$ inertia can be ignored and the system is effectively overdamped. After neglecting the inertial term $m\ddot{x}$ in Eq.~(\ref{trap}) the equation of motion in the overdamped regime can then be written as
\begin{eqnarray}
\label{OD}
\dot{x}(t)=-\frac{1}{\tau_r}(x(t)-vt)+\frac{1}{\alpha}\xi(t),
\end{eqnarray}
or, respectively, in the comoving frame
\begin{eqnarray}
\label{OD_y}
\dot{y}(t)=-\frac{1}{\tau_r} y(t)-\left(v+\frac{\lambda\Gamma_0}{\alpha}\right)+\frac{1}{\alpha}z(t),
\end{eqnarray}
where we have used Eq.~(\ref{xi}). We see that the subtracted mean value of Eq.~(\ref{xi}) has the effect of an additional drift force on the particle in Eq.~(\ref{OD_y}). Introducing an effective velocity $v_e$ defined as
\begin{eqnarray}
\label{v_eff}
v_{e}\equiv v+\lambda\Gamma_0/\alpha,
\end{eqnarray}
allows us to write Eq.~(\ref{OD_y}) as
\begin{eqnarray}
\label{F_y}
\dot{y}(t)=-\frac{1}{\tau_r} y(t)-v_{e}+\frac{1}{\alpha}z(t),
\end{eqnarray}
The one-sidedness of the shot noise $z(t)$ leads to singular features of the work fluctuations in the overdamped regime, as we discuss in more detail in the following.

\subsection{Singularities in the overdamped regime}
\label{Sec_sing_OD}

From the Langevin equation~(\ref{F_y}) we can infer two important properties of the model in the overdamped regime. Firstly, upon averaging of Eq.~(\ref{F_y}) we obtain
\begin{eqnarray}
\label{mean_eq}
\frac{\upd}{\upd t}\left<y(t)\right>=-\frac{1}{\tau_r}\left<y(t)\right>-v.
\end{eqnarray}
In the NESS the time derivative on the left hand side (lhs) is zero and the stationary mean position $\left<y\right>$ is simply given by
\begin{eqnarray}
\label{mean_pos}
\left<y\right>=-v\tau_r.
\end{eqnarray}
Then, using Eq.~(\ref{work_par}), we find the mean value of the work in the steady state
\begin{eqnarray}
\label{mean_work}
\left<W_\tau\right>&=&-v\kappa\int_0^{\tau} \left<y\right>\upd t=\alpha v^2\tau.
\end{eqnarray}
Clearly, $\left<W_\tau\right>$ is always positive as required by the second law. These expressions for the mean position and mean work are the same as in the Gaussian case (see e.g. \cite{VanZon03}) due to the zero mean of the noise in both cases.

\begin{figure}
\begin{center}
\includegraphics[height=4cm]{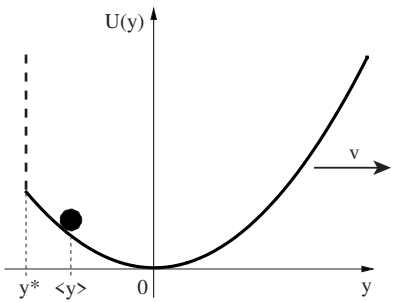}
\caption{\label{Fig_nonlinear1}The truncated (effective) harmonic potential $U(y)$ in the comoving frame for $v>0$ (regime (i)). The black bullet depicts the particle at its mean position $\left<y\right>$, while the effective infinite barrier for the particle position at $y^*$ is indicated with a dashed black line. For $v>0$ both the mean position $\left<y\right>$ and the cut-off $y^*$ are always negative (cf. Eqs.~(\ref{mean_pos}) and (\ref{min_pos})).  }
\end{center}
\end{figure}

Secondly, we find that there exists a \textit{minimal value} $y^*$ of the position coordinate. This can be understood if we consider Eq.~(\ref{F_y}) without $z(t)$, that is
\begin{eqnarray}
\label{min_pos_z_eq}
\dot{y}(t)=-\frac{1}{\tau_r} y(t)-v_{e}.
\end{eqnarray}
In the NESS, where the lhs of Eq.~(\ref{min_pos_z_eq}) is zero, the particle will reach a position $y^*$ given by
\begin{eqnarray}
\label{min_pos}
y^*&=&-v_{e}\tau_r.
\end{eqnarray}
Since the shot noise part $z(t)$ can only move the particle in the positive $y$-direction, this position $y^*$ is the minimal position the particle can reach, i.e., $y^*$ represents a \textit{cut-off} in position space. The origin of this cut-off is the asymmetric form of the noise $\xi(t)$, Eq.~(\ref{xi}), which is given as a superposition of a shot noise part $z(t)$ acting in the positive direction only and of a drift part acting in the negative direction only. Without $z(t)$ the particle will move until it reaches the position $y^*$ where the negative drift force is balanced by the positive spring force. Under the influence of $z(t)$ the particle can then only reach positions to the right of $y^*$ (cf. Fig.~\ref{Fig_nonlinear1}).

The presence of a position cut-off implies that the parabolic potential becomes effectively \textit{nonlinear}: the effect of $y^*$ is that of an infinite barrier in the potential. The effective potential for the particle is given by the harmonic potential $U(y)$, truncated at $y^*$ (cf. Fig.~\ref{Fig_nonlinear1}). Since the effective nonlinearity arises due to the asymmetry of the PSN, our model exhibits an effective interaction between the noise and the potential.

From Eq.~(\ref{work_par}) we find that the work rate $w(t)$ is given by
\begin{eqnarray}
\label{work_rate}
w(t)=-v\kappa y(t),
\end{eqnarray}
i.e., proportional to the position. The position cut-off therefore implies a cut-off of the work rate $w(t)$, so that we also obtain a \textit{cut-off value of the work} in time $\tau$ in the NESS, namely
\begin{eqnarray}
\label{ex_work}
W^*_\tau&=&-v\kappa\int_0^\tau y^*\upd t=- v\kappa y^*\tau=\alpha v v_e\tau.
\end{eqnarray}
We can also express the position and work cut-offs in terms of the time scales $\tau_\lambda$ and $\tau_p$. Using Eqs~(\ref{tau_l}) and~(\ref{tau_p}) allows us to write $y^*$ of Eq.~(\ref{min_pos}) in the form
\begin{eqnarray}
\label{pos_tp}
y^*=-v\left(1+\sigma(v)\frac{\tau_p}{\tau_\lambda}\right)\tau_r,
\end{eqnarray}
where $\sigma(v)$ denotes the sign function defined as
\begin{eqnarray}
\sigma(v)=\left\{\begin{array}{c} +1\quad, \qquad v>0\\ -1 \quad, \qquad v<0\\ 0 \quad\quad,\qquad v=0 . 
\end{array}\right.
\end{eqnarray}
The work cut-off of Eq.~(\ref{ex_work}) then reads
\begin{eqnarray}
\label{work_tp}
W^*_\tau=\left<W_\tau\right>\left(1+\sigma(v)\frac{\tau_p}{\tau_\lambda}\right).
\end{eqnarray}

Due to $\Gamma_0>0$ the effective velocity, Eq.~(\ref{v_eff}), obeys always $v_e>v$ so that $y^*<\left<y\right>$, i.e., the cut-off is always to the left of the mean position. Both work and position cut-offs have different characteristics depending on the sign of $v$ and the ratio of $\tau_p$ and $\tau_r$. We distinguish three different regimes which are important for the later discussion of the work fluctuations (Sec.~\ref{Sec_FT}).

(i) $v>0$ (see Fig.~\ref{Fig_nonlinear1}). In this case the mean position is $\left<y\right><0$ and the cut-off $y^*$ of Eq.~(\ref{pos_tp}) is also always $<0$. The work cut-off $W^*_\tau$ is then $>0$ (cf. Eq.~(\ref{work_tp})) and denotes the \textit{maximal} work done on the particle in time $\tau$. This is due to the fact that for $v>0$ the work rate $w(t)$, Eq.~(\ref{work_rate}), is larger for smaller positions $y(t)$.

\begin{figure}
\begin{center}
\includegraphics[height=4cm]{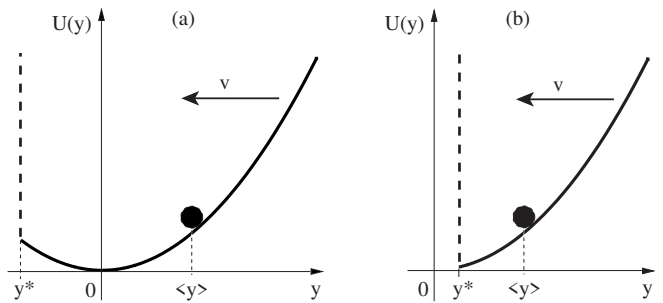} 
\caption{\label{Fig_nonlinear2}The effective harmonic potential in the comoving frame for $v<0$. Here, the mean position $\left<y\right>$ is positive (Eq.~(\ref{mean_pos})) while the cut-off $y^*$ can be positive or negative (cf. Eq.~(\ref{pos_tp})). (a) Negative $y^*$, if $\tau_p>\tau_\lambda$ (regime (ii)). (b) Positive $y^*$, if $\tau_p<\tau_\lambda$ (regime (iii)). }
\end{center}
\end{figure}

(ii) $v<0$ and $\tau_p>\tau_\lambda$ (see Fig.~\ref{Fig_nonlinear2}a). Both $y^*$ and $W^*_\tau$ are then negative (cf. Eqs.~(\ref{pos_tp}) and (\ref{work_tp})). Moreover, the work cut-off is now the \textit{minimal} work performed over time $\tau$, since for negative $v$ the work rate $w(t)$ of Eq.~(\ref{work_rate}) is smaller for smaller positions.

(iii) $v<0$ and $\tau_p<\tau_\lambda$ (see Fig.~\ref{Fig_nonlinear2}b). In this case both $y^*$ and $W^*_\tau$ are positive. As in (ii), $W^*_\tau$ is the minimal work performed over time $\tau$. However, a positive minimal work value $W^*_\tau$ implies that no negative work fluctuations can occur.

Due to the existence of the cut-off values $y^*$ and $W^*_\tau$ the distributions of both position and work are non-Gaussian, unless one considers the Gaussian limit of the PSN Eqs.~(\ref{G_limit})---(\ref{G_limit_th}). In that limit $y^*\rightarrow -\infty$ and $W^*_\tau\rightarrow\pm\infty$. We remark that a similar work cut-off has been first observed in a Brownian particle model, where the moving potential is given as a nonlinear potential of the Lennard-Jones-type \cite{Dykman}. 

The position and work cut-off are properties of the overdamped regime only. If the dynamics is influenced by inertia the cut-offs disappear, because the particle can actually `overshoot' the minimal position $y^*$, following a strong fluctuation of $z(t)$ in the positive direction, due to its mass. However, this inertial effect is only relevant on short time scales and disappears if $\tau\gg\tau_m$. In the asymptotic time regime therefore, in which we are mainly interested here, the fluctuation properties of the particle are described by the overdamped equation of motion and exhibit the singular features discussed above.

\section{Calculation of the characteristic function of the work for PSN}
\label{Sec_cf}

An elegant method to calculate the probability distribution of $W_\tau$ has been presented in \cite{Touchette07,Touchette09}. This calculation is based on a theorem for a generalized Ornstein-Uhlenbeck process \cite{Caceres97}, which is also applicable in the present case, if we write the equation of motion (\ref{oscillator2}) in terms of a two-component system for the position $y(t)$ and velocity $u(t)$ of the particle in the comoving frame:
\begin{eqnarray}
\dot{y}(t)&=&u(t),\\
\dot{u}(t)&=&-\frac{1}{\tau_m}u(t)-\frac{1}{\tau_m\tau_r}y(t)-\frac{1}{\tau_m}v_e+\frac{1}{m}z(t),
\end{eqnarray}
using Eqs.~(\ref{v_eff}) and (\ref{xi}). In vector-matrix notation this equation can be expressed as a linear first order differential equation of the form
\begin{eqnarray}
\label{matrix_eq}
\dot{\mathbf{y}}(t)=\mathcal{M}\,\mathbf{y}(t)+\mathbf{a}+\mathbf{z}(t),
\end{eqnarray}
where we have defined the vectors
\begin{eqnarray}
\label{vectors}
\mathbf{y}(t)&\equiv& \left(\begin{array}{c}
y(t) \\ u(t)
\end{array}\right),\qquad \mathbf{a}\equiv \left(\begin{array}{c}
0 \\  -\frac{1}{\tau_m}v_e
\end{array}\right), \nonumber\\
 \mathbf{z}(t)&\equiv&\left(\begin{array}{c}
0 \\ \frac{1}{m}z(t)
\end{array}\right),
\end{eqnarray}
while the matrix $\mathcal{M}$ is defined as
\begin{eqnarray}
\mathcal{M}\equiv\left(\begin{array}{cc}
0 & 1 \\ -\frac{1}{\tau_m\tau_r} & -\frac{1}{\tau_m}
\end{array}\right).
\end{eqnarray}
The derivation of the work distribution then proceeds along the following steps:

1. The stochastic process $\mathbf{y}(t)$ of Eq.~(\ref{matrix_eq}) describes a generalized (two-component) Ornstein-Uhlenbeck process \cite{VanKampen}. In order to determine the characteristic functional of $\mathbf{y}(t)$, defined by
\begin{eqnarray}
\label{cf_def}
G_{\mathbf{y}(t)}[\mathbf{h}(t)]\equiv\left<\exp\left\{i\int_0^\infty \mathbf{h}(t)\mathbf{y}(t)\upd t\right\}\right>,
\end{eqnarray}
for a two-component test function $\mathbf{h}(t)$, we can then apply the two-component version of the C\'aceres-Budini theorem \cite{Caceres97} (see Appendix~\ref{App_cb}). This theorem states that $G_{\mathbf{y}(t)}[\mathbf{h}(t)]$ follows from the characteristic noise functional $G_{\mathbf{z}(t)}[\mathbf{k}(t)]$ via:
\begin{eqnarray}
\label{cb}
G_{\mathbf{y}(t)}[\mathbf{h}(t)]=e^{i\mathbf{y}_0\mathbf{k}_0+i\mathbf{a}\int_0^\infty\mathbf{k}(t)\upd t}G_{\mathbf{z}(t)}[\mathbf{k}(t)],
\end{eqnarray}
where $\mathbf{k}(t)$ is given by
\begin{eqnarray}
\label{cb_kt}
\mathbf{k}(t)&=&\left(\begin{array}{c}
k_1(t) \\  k_2(t) \end{array}\right)\equiv\int_t^\infty e^{\mathcal{M}^{\rm T}(s-t)}\mathbf{h}(s)\upd s
\end{eqnarray}
and
\begin{eqnarray}
\mathbf{k}_0 \equiv\mathbf{k}(t=0)=\left(\begin{array}{c}
k_1(0) \\  k_2(0) \end{array}\right).
\end{eqnarray}
The initial position $y_0$ and initial velocity $u_0$ of the particle are contained in $\mathbf{y}_0\equiv(y_0,u_0)^{\rm T}$. Due to the zero first component of $\mathbf{z}(t)$ (cf. Eq.~(\ref{vectors})), the noise functional in Eq.~(\ref{cb}) is given by $G_{\mathbf{z}(t)}[\mathbf{k}(t)]=G_{z(t)}[k_2(t)/m]$, where $G_{z(t)}[g(t)]$, the characteristic functional of the PSN $z(t)$, has the exact form Eq.~(\ref{noise_functional}).

2. We obtain the characteristic function of the work
\begin{eqnarray}
G_{W_\tau}(q)\equiv\left<e^{iqW_\tau}\right>
\end{eqnarray}
by considering the particular test function (cf. \cite{Touchette07,Touchette09})
\begin{eqnarray}
\label{hugo_k}
\mathbf{\tilde{h}}(t)=\left(\begin{array}{c}
 - q v \kappa \Theta(\tau-t) \\ 0
\end{array}\right),
\end{eqnarray}
in the characteristic functional $G_{\mathbf{y}(t)}[\mathbf{h}(t)]$, where $\Theta(t)$ denotes the Heaviside step function. This can be seen immediately by substituting Eq.~(\ref{hugo_k}) for $\mathbf{h}(t)$ into the definition of $G_{\mathbf{y}(t)}[\mathbf{h}(t)]$, Eq.~(\ref{cf_def}), i.e.
\begin{eqnarray}
G_{\mathbf{y}(t)}\left[\mathbf{\tilde{h}}(t)\right]=G_{W_\tau}(q),
\end{eqnarray}
due to the expression for the work $W_\tau$ Eq.~(\ref{work_par}).

3. Finally, the work distribution follows by performing the inverse Fourier transform of $G_{W_\tau}(q)$. In the overdamped regime this can be done analytically using the method of steepest descent (see Sec.~\ref{Sec_FT}).

We now evaluate the functional $\mathbf{k}(t)$ using Eq.~(\ref{cb_kt}). The matrix exponential $\exp\left\{{\mathcal{M}}^{\rm T}\right\}$ can be determined by diagonalization of the matrix $\mathcal{M}$. Using the particular test function $\mathbf{\tilde{h}}(t)$ of Eq.~(\ref{hugo_k}) one obtains then the components of $\mathbf{k}(t)$ in a straightforward way
\begin{eqnarray}
\label{k_component1}
k_1(t)&=&\frac{q\,v\,\kappa}{\nu_1-\nu_2}\Theta(\tau-t)\left(\frac{\nu_1}{\nu_2}\left(1-e^{\nu_2(\tau-t)}\right)\right.\nonumber\\
&&\left.-\frac{\nu_2}{\nu_1}\left(1-e^{\nu_1(\tau-t)}\right)\right),\\
\label{k_component2}
k_2(t)&=&\frac{q\,v\,\kappa}{\nu_1-\nu_2}\Theta(\tau-t)\left(\frac{1}{\nu_1}\left(1-e^{\nu_1(\tau-t)}\right)\right.\nonumber\\
&&\left.-\frac{1}{\nu_2}\left(1-e^{\nu_2(\tau-t)}\right)\right).
\end{eqnarray}

Substituting these expressions into Eq.~(\ref{cb}) leads to an explicit expression for the characteristic function of the work:
\begin{widetext}
\begin{eqnarray}
\label{cf_work}
G_{W_\tau}(q)&=&\exp\left\{i\mathbf{y}_0\mathbf{k}_0-iq\frac{v\,v_e\,\kappa}{\tau_m(\nu_1-\nu_2)}
\left(\frac{1}{\nu_1^2}\left(\nu_1\tau+1-e^{\nu_1\tau}\right)-\frac{1}{\nu_2^2}\left(\nu_2\tau+1-e^{\nu_2\tau}\right)\right)\right.\nonumber\\
&&\left.+\frac{1}{\tau_\lambda}\int_0^\tau\left(\frac{1}{1-iq \Gamma_0\frac{v\,\kappa}{m(\nu_1-\nu_2)}\left(\frac{1}{\nu_1}\left(1-e^{\nu_1(\tau-t)}\right)-\frac{1}{\nu_2}\left(1-e^{\nu_2(\tau-t)}\right)\right)}-1\right)\upd t\right\},
\end{eqnarray}
\end{widetext}
where $\mathbf{y}_0$ contains the initial conditions $y_0$ and $u_0$ and the components of $\mathbf{k}_0$ are obtained from Eqs.~(\ref{k_component1}) and (\ref{k_component2}) by setting $t=0$. The work distribution now follows by performing the inverse Fourier-transform of Eq.~(\ref{cf_work}), which we have not been able to perform exactly. Nevertheless, Eq.~(\ref{cf_work}) can be readily used in order to determine the hierarchy of cumulants for the work, which reveal an oscillatory behavior if the influence of inertia is non-negligible. This is further studied in Sec.~\ref{Sec_inertia}. In the next section we focus on the overdamped regime, where an analytical form of the work distribution can be obtained.  

\section{Asymptotic fluctuation properties}
\label{Sec_FT}

\begin{figure}
\begin{center}
\includegraphics[width=7cm]{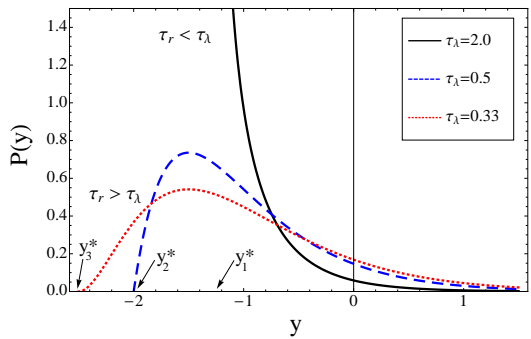}
\caption{\label{Fig_dist_yP}The distribution of the particle position in the NESS, $P(y)$ of Eq.~(\ref{p_stat}). One observes the divergence for $y\rightarrow y^*$ if $\tau_r<\tau_\lambda$. The cut-off values are given by $y_1^*=-1.25$, $y_2^*=-2.0$, and $y_3^*=-2.5$. Parameter values: $\tau_r=1.0$, $v=1.0$, $\Gamma_0=0.5$ \cite{Footnote3}.}
\end{center}
\end{figure}

In the asymptotic regime $\tau\rightarrow\infty$ inertial effects can be neglected and the work fluctuations behave as in the overdamped system. The characteristic function of the work in the overdamped regime is obtained by taking the limit $m\rightarrow 0$ in Eq.~(\ref{cf_work}), when the eigenvalues Eq.~(\ref{eigenvalues}) are given by $\nu_1\approx -\tau_r^{-1}$ and $\nu_2\approx -\tau_m^{-1}$, respectively. The characteristic function of the work Eq.~(\ref{cf_work}) then becomes
\begin{widetext}
\begin{eqnarray}
\label{cf_work_od}
G_{W_\tau}(q)&=&\exp\left\{iy_0k_0+iqW_\tau^*
\left(1-\left(1-e^{-\tau/\tau_r}\right)\tau_r/\tau\right)+\frac{1}{\tau_\lambda}\int_0^\tau\left(\frac{1}{1+iq \Gamma_0 v\left(1-e^{(t-\tau)/\tau_r}\right)}-1\right)\upd t\right\},
\end{eqnarray}
where $y_0$ is the initial position and $k_0$ denotes the upper component of $\mathbf{k}_0$ in the overdamped regime, which reads $k_0=-q\alpha v\left(1-e^{-\tau/\tau_r}\right)$. The integral in Eq.~(\ref{cf_work_od}) can be expressed in closed form leading to an exact expression for the characteristic function of the work:
\begin{eqnarray}
\label{F_cf_poiss_exp}
G_{W_\tau}(q)&=&\left(1+iq\Gamma_0 v(1-e^{-\tau/\tau_r})\right)^\frac{\tau_r/\tau_\lambda}{1+iq \Gamma_0 v }\exp\left\{iy_0k_0+iqW_\tau^*\left(1-\left(1- e^{-\tau/\tau_r}\right)\frac{\tau_r}{\tau}\right)+\frac{\tau}{\tau_\lambda}\left(\frac{1}{1+iq \Gamma_0 v }-1\right)\right\}.
\end{eqnarray}
\end{widetext}

This expression simplifies if we consider the NESS of the system, where the initial position $y_0$ is drawn from the stationary nonequilibrium distribution. This distribution can be found by solving the Fokker-Planck equation associated with the Langevin equation~(\ref{F_y}). We obtain then for the distribution of the particle position in the NESS (see Appendix~\ref{App_ss})
\begin{eqnarray}
\label{p_stat}
P(y)=\frac{1}{\Gamma(\frac{\tau_r}{\tau_\lambda})}\frac{\alpha}{\Gamma_0}\left(\frac{\alpha}{\Gamma_0}(y-y^*)\right)^{\frac{\tau_r}{\tau_\lambda}-1} e^{-(y-y^*)\alpha/\Gamma_0},
\end{eqnarray}
where $\Gamma(\tau_r/\tau_\lambda)$ denotes the Gamma function \cite{Abramowitz} with argument $\tau_r/\tau_\lambda$. One notices that the exponent in the prefactor of the exponential in Eq.~(\ref{p_stat}) becomes negative if $\tau_r<\tau_\lambda$, so that then $P(y)$ diverges for $y\rightarrow y^*$. This singular behavior is related physically to insufficient noise activation in the system when $\tau_r<\tau_\lambda$. For, the time $\tau_\lambda$ is the average waiting time between two successive pulses of the shot noise. Therefore, if $\tau_r<\tau_\lambda$ the system relaxes fast compared to the time scale of noise activation. In other words, the system relaxes `too quickly' in between stochastic pulses and thus spends most of its time at the position that it would assume deterministically without the shot noise, which is $y^*$. Consequently $P(y)$ diverges for $y\rightarrow y^*$ (cf. Fig.~\ref{Fig_dist_yP}). 

We can now average the initial position $y_0$ in the expression for the characteristic function, Eq.~(\ref{F_cf_poiss_exp}), over the stationary distribution Eq.~(\ref{p_stat}). The result reads
\begin{eqnarray}
\label{cf_final}
G_{W_\tau}(q)&=&\left(1+iq\Gamma_0v(1-e^{-\tau/\tau_r})\right)^{\frac{\tau_r}{\tau_\lambda}\left(\frac{1}{1+iq \Gamma_0 v}-1\right)}\nonumber\\&&\exp\left\{iqW_\tau^*+\frac{\tau}{\tau_\lambda}\left(\frac{1}{1+iq \Gamma_0 v }-1\right)\right\},
\end{eqnarray}
where $W^*_\tau$ is given by Eq.~(\ref{ex_work}). The work distribution is obtained by carrying out an inverse Fourier transform of Eq.~(\ref{cf_final}) and allows then an investigation of the work fluctuation properties in the asymptotic regime.

\subsection{Large deviation form of the work distribution}
\label{Sec_large_Dev}

The inverse Fourier transform of Eq.~(\ref{cf_final}) can be calculated analytically using the method of steepest descent. It is convenient to introduce the scaled dimensionless value of the work $p$, defined by
\begin{eqnarray}
p\equiv\frac{W_\tau}{\left<W_\tau\right>}.
\end{eqnarray}
The work cut-off $W^*_\tau$, Eq.~(\ref{work_tp}), then gives rise to an extremal value $p^*$ of the scaled work $p$, defined by
\begin{eqnarray}
\label{p_cutoff}
p^*\equiv\frac{W_\tau^*}{\left<W_\tau\right>}=1+\sigma(v)\frac{\tau_p}{\tau_\lambda}.
\end{eqnarray}
This expression implies that the three regimes of the work fluctuations, discussed quantitatively in Sec.~\ref{Sec_sym} (below Eq.~(\ref{work_tp})), correspond to three different regimes of $p^*$, namely: (i) $v>0$ implies $p^*>1$. (ii) $v<0$ and $\tau_p > \tau_\lambda$ implies $p^*<0$. (iii) $v<0$ and $\tau_p < \tau_\lambda$ implies $0<p^*<1$.

The distribution function of $p$, denoted by $\Pi_\tau(p)$, is obtained from the inverse Fourier-transform of $G_{W_\tau}$, Eq.~(\ref{cf_final}) by
\begin{eqnarray}
\label{inverseFourier}
\Pi_\tau(p)=\frac{\left<W_\tau\right>}{2\pi}\int_{-\infty}^\infty G_{W_\tau}(q)e^{-iqp\left<W_\tau\right>}\upd q.
\end{eqnarray}
Using Eq.~(\ref{cf_final}) on the rhs of Eq.~(\ref{inverseFourier}) as well as Eq.~(\ref{mean_work}) and Eq.~(\ref{ex_work}), we see that the distribution $\Pi_\tau(p)$ can be written in the form
\begin{eqnarray}
\label{FT_int}
\Pi_\tau(p)=\frac{\alpha v^2\tau}{2\pi}\int_{-\infty}^\infty\chi(q)e^{\tau h(q)}\upd q,
\end{eqnarray}
where, the functions $\chi(q)$ and $h(q)$ are given by
\begin{eqnarray}
\chi(q)\equiv\left(1+iq\Gamma_0v\right)^{\frac{\tau_r}{\tau_\lambda}\left(\frac{1}{1+iq \Gamma_0 v}-1\right)},
\end{eqnarray}
and
\begin{eqnarray}
\label{sp_h}
h(q)\equiv iq\alpha v^2 (p^*-p)+\frac{1}{\tau_\lambda}\left(\frac{1}{1+iq \Gamma_0 v }-1\right),
\end{eqnarray}
respectively.

For large $\tau$ the integral in Eq.~(\ref{FT_int}) will be dominated by its saddle-point and can be approximately calculated using the method of steepest descent \cite{Jeffreys}. Neglecting terms of order $\tau^{-1/2}$ leads then to the saddle-point approximation of $\Pi_\tau(p)$ in the form
\begin{eqnarray}
\label{sp_dist}
\Pi_\tau(p)\cong\frac{\alpha v^2}{\sqrt{2\pi}}\sqrt{\frac{\tau}{|h''(\bar{q})|}}\chi(\bar{q})e^{i\theta+\tau h(\bar{q})},
\end{eqnarray}
where $\bar{q}$ denotes the appropriate saddle-point and $\theta$ the angle between the deformed integration path and the real axis. Saddle-points are determined from the condition $h'(\bar{q})=0$, which, with Eq.~(\ref{sp_h}), reads 
\begin{eqnarray}
(1+i\bar{q} \Gamma_0 v )^2-\frac{\Gamma_0}{\alpha v \tau_\lambda(p^*-p)}=0.
\end{eqnarray}
The solution of this quadratic equation is given by
\begin{eqnarray}
\bar{q}_{\pm}=\frac{i}{\Gamma_0v}\left(1\pm\sqrt{\frac{p^*-1}{p^*-p}}\right).
\end{eqnarray}
Importantly, the square root $\sqrt{(p^*-p)/(p^*-1)}$ is always real, since both $p^*-p$ and $p^*-1$ are either both positive ($v>0$) or negative ($v<0$), respectively. This means that the real part $\mathcal{R}(\bar{q}_\pm)=0$ for all $p$ and we can conclude that the appropriate saddle-point is
\begin{eqnarray}
\label{saddle_point_correct}
\bar{q}_-=\frac{i}{\Gamma_0v}\left(1-\sqrt{\frac{p^*-1}{p^*-p}}\right),
\end{eqnarray}
since the original integration path (the real axis) can be deformed to go through $\bar{q}_-$ without crossing the pole at $q=i/(\Gamma_0 v)$. Furthermore $\theta=0$ in Eq.~(\ref{sp_dist}), as is required for a real probability distribution, since $h(q)$ is real for purely imaginary $q$ (cf. Eq.~(\ref{sp_h})) and therefore the path of steepest descent through $\bar{q}_-$ is parallel to the real axis (cf. the discussion in Sec.~IV of \cite{VanZon04}).

Substituting the appropriate saddle-point Eq.~(\ref{saddle_point_correct}) into the expressions for $\chi(q)$, $h(q)$, and $h''(q)$ yields for the saddle-point approximation Eq.~(\ref{sp_dist}):
\begin{eqnarray}
\label{p_dist}
\Pi_\tau(p)&\cong&\frac{1}{\sqrt{4\pi}}\frac{\sqrt{\tau/\tau_\lambda}}{|p^*-1|}\left(\sqrt{\frac{p^*-p}{p^*-1}}\right)^{-\frac{\tau_r}{\tau_\lambda}\left(\sqrt{\frac{p^*-p}{p^*-1}}-1\right)-\frac{3}{2}}\nonumber\\
&&\exp\left\{-\frac{\tau}{\tau_\lambda}\left(\sqrt{\frac{p^*-p}{p^*-1}}-1\right)^2\right\}.
\end{eqnarray}
With Eq.~(\ref{p_cutoff}) we see that the distribution $\Pi_\tau(p)$ is completely specified by the time scales $\tau_r$, $\tau_\lambda$, $\tau_p$, in addition to the measurement time $\tau$.

One notices two different singularities appearing in Eq.~(\ref{p_dist}). Firstly, the derivative of $\Pi_\tau(p)$ diverges for $p\rightarrow p^*$ as $\Pi'(p)\propto|p^*-p|^{-1}$ in leading order. This means that the approach of $\Pi_\tau(p)$ to the cut-off $p^*$ has a vertical slope. Secondly, one notices that $\Pi_\tau(p)$ itself diverges for $p\rightarrow p^*$ if $\tau_r/\tau_\lambda<3/2$. However, for large $\tau$ this divergence occurs only in an isolated point and can be ignored.

\begin{figure}
\begin{center}
\includegraphics[width=8cm]{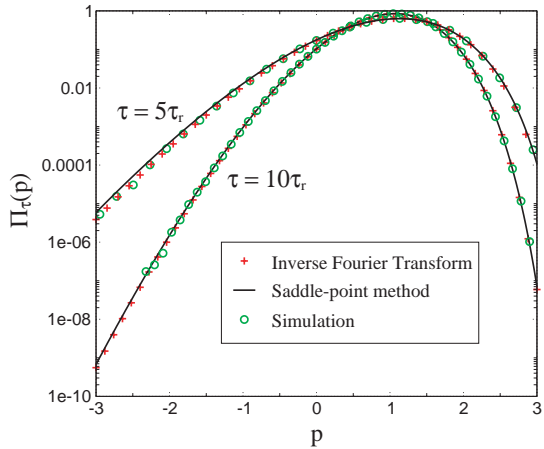}
\caption{\label{Fig_logdist}Comparison of the analytic saddle-point approximation for $\Pi_\tau(p)$, Eq.~(\ref{p_dist}), with a numerical inverse Fourier transform of $G_{W_\tau}$ as well as with results from a direct simulation of the equation of motion (\ref{F_y}). Parameter values: $v=1$, $\tau_r=1$, $\tau_\lambda=0.2$, $\tau_p=0.5$.}
\end{center}
\end{figure}

We find that Eq.~(\ref{p_dist}) yields an excellent approximation of the distribution $\Pi_\tau(p)$ at least for $\tau\geq 10\tau_r$. This is shown in Fig.~\ref{Fig_logdist}  where we compare Eq.~(\ref{p_dist}) with a numerical inverse Fourier transform of $G_{W_\tau}$ and also with results from a direct simulation of the equation of motion (\ref{F_y}) using a Poissonian increment method \cite{Kim07}. For $\tau=5\tau_r$ one notices a slight deviation in the negative tail for $p<-1$ between the saddle-point result and the numerical inverse Fourier transform. For $\tau=10\tau_r$ the saddle-point approximation is in excellent agreement with the inverse Fourier transform over the whole range of $p$ values.

\begin{figure}
\begin{center}
\includegraphics[width=7cm]{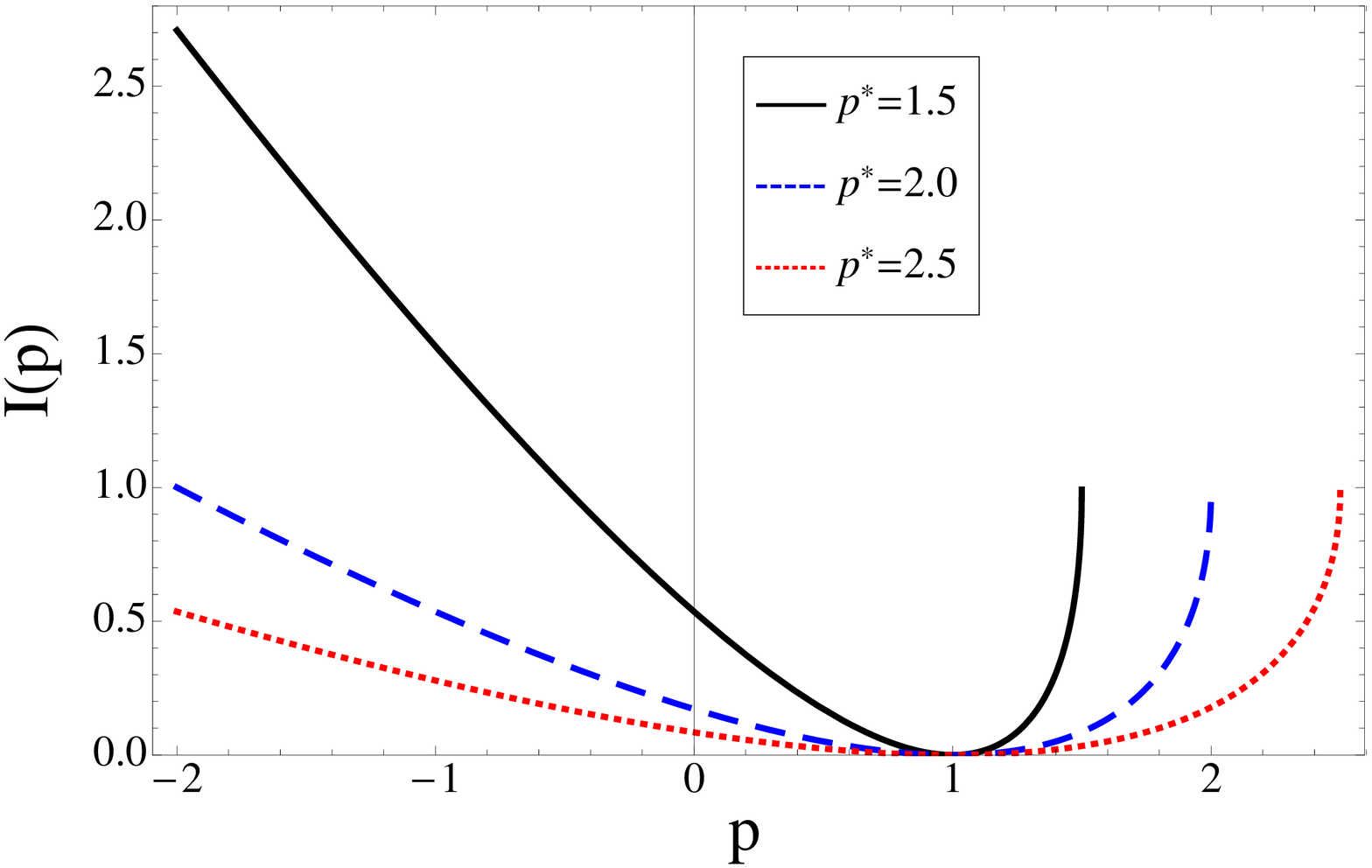}
\caption{\label{Fig_rf_p1}The rate function $I(p)$, Eq.~(\ref{rate_func}), for $v>0$ and $\tau_\lambda=1$, plotted as a function of $p$ for three different $p^*$ values. At every $p^*$, $I(p)$ assumes the finite value $I(p^*)=1/\tau_\lambda=1$ here, which is approached with a vertical slope. For large negative $p$, $I(p)$ becomes asymptotically linear. All curves correspond to regime (i).}
\end{center}

\begin{center}
\includegraphics[width=7cm]{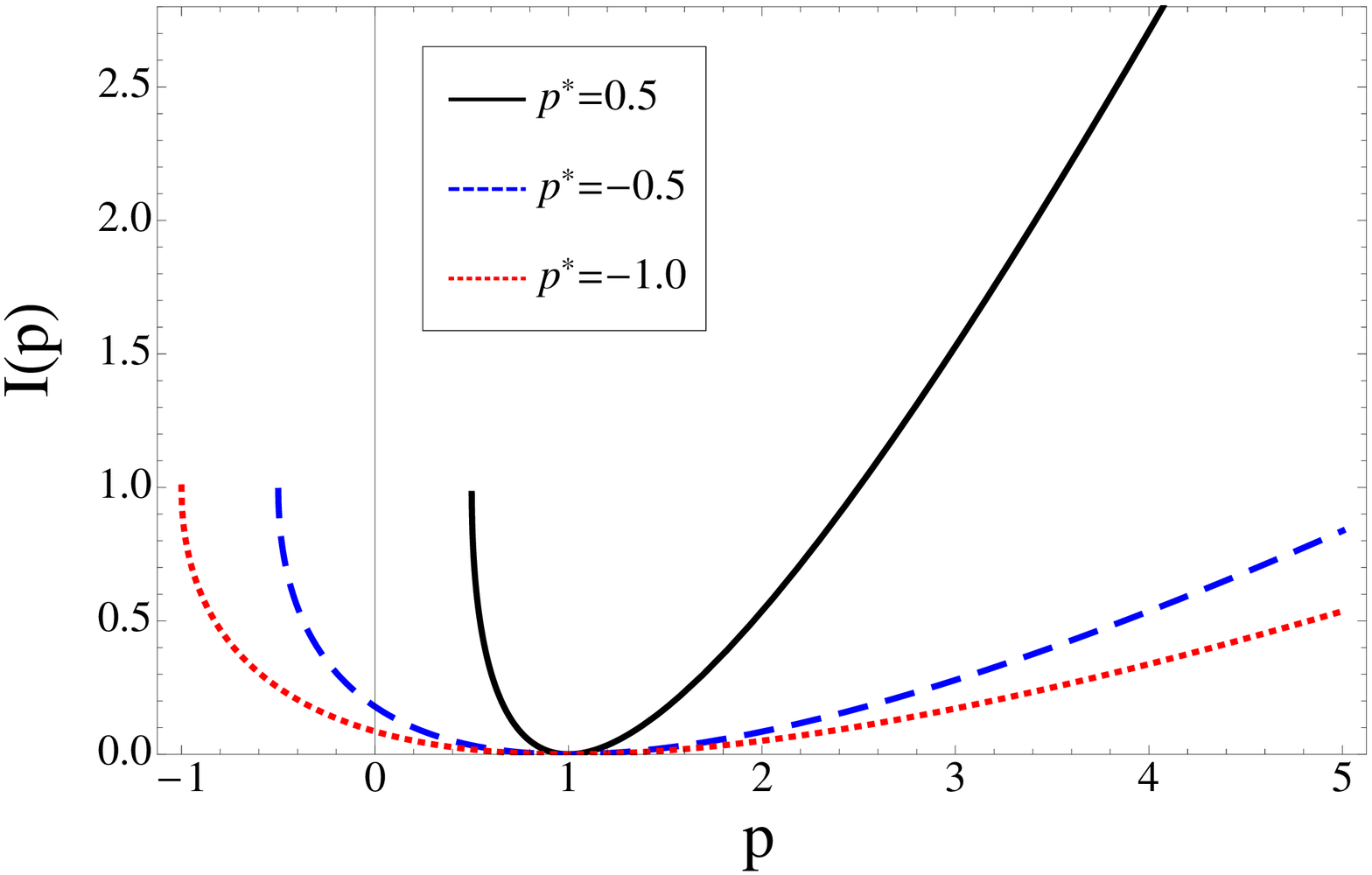}
\caption{\label{Fig_rf_p2}The rate function $I(p)$ of Eq.~(\ref{rate_func}) for $v<0$ and $\tau_\lambda=1$, plotted as a function of $p$ for three different $p^*$ values. At every $p^*$ the rate function ends at the finite value $I(p^*)=1/\tau_\lambda=1$ here, which is approached with a vertical slope. For large positive $p$, $I(p)$ becomes asymptotically linear. The dashed and dotted curves correspond to regime (ii), where $p^*<0$. The solid curve corresponds to regime (iii), where $0<p^*<1$ and no negative work fluctuations occur.}
\end{center}

\end{figure}

For very large $\tau$ the distribution $\Pi_\tau(p)$ exhibits the large deviation form \cite{Touchette08}
\begin{eqnarray}
\label{largeDev}
\Pi_\tau(p)\cong e^{-\tau I(p)}
\end{eqnarray}
with rate function
\begin{eqnarray}
\label{rate_func}
I(p)\equiv\frac{1}{\tau_\lambda}\left(\sqrt{\frac{p^*-p}{p^*-1}}-1\right)^2.
\end{eqnarray}
The rate function $I(p)$ of Eq.~(\ref{rate_func}) is plotted separately for $v>0$ and $v<0$ in Figs.~\ref{Fig_rf_p1} and \ref{Fig_rf_p2}, respectively. In both cases one observes the strongly asymmetric shape of the rate function, which attains its minimum at the most likely work value, namely at $p=1$, i.e., at the mean work value as expected.

(i) $v>0$ (see Fig.~\ref{Fig_rf_p1}). The work cut-off is here $p^*>1$ and denotes the maximal work done on the particle. As $p\rightarrow p^*$ the rate function $I(p)$ approaches $I(p^*)$ with a vertical slope and ends at the finite value $I(p^*)=1/\tau_\lambda$ (cf. Eq.~(\ref{rate_func})). The left side of $I(p)$ is unbounded and becomes asymptotically linear for large negative $p$. From Eq.~(\ref{largeDev}) it follows that the work distribution $\Pi_\tau(p)$ decays exponentially for large negative $p$.

(ii) $v<0$ and $\tau_p>\tau_\lambda$ (see dashed and dotted curves in Fig.~\ref{Fig_rf_p2}). The work cut off is now $p^*<0$ and denotes the minimal work done on the particle. $I(p)$ approaches the end point $I(p^*)=1/\tau_\lambda$ with a vertical slope as $p\rightarrow p^*$. Now, the right side of $I(p)$ is unbounded and becomes asymptotically linear for large $p$, so that $\Pi_\tau(p)$ decays exponentially for large positive $p$.

(iii) $v<0$ and $\tau_p<\tau_\lambda$. Here, the work cut off is $0<p^*<1$ and no negative work fluctuations occur as becomes evident in Fig.~\ref{Fig_rf_p2} (solid curve). The behavior in the approach to the cut-off and for large $p$ is as in (ii).

\subsection{Fluctuation theorem}
\label{Sec_poiss_FT}

In the asymptotic regime $\tau\rightarrow \infty$ the work distribution Eq.~(\ref{p_dist}) has the large deviation form Eq.~(\ref{largeDev}). In order to further discuss the fluctuation properties of work it is convenient to consider the dimensionless fluctuation function
\begin{eqnarray}
\label{fluctuations}
f_\tau(p)\equiv \frac{1}{a\left<W_\tau\right>}\ln\frac{\Pi_\tau(p)}{\Pi_\tau(-p)},
\end{eqnarray}
where the constant $a$ is defined by
\begin{eqnarray}
\label{const_a}
a\equiv\frac{\alpha\tau_\lambda}{\Gamma^2_0},
\end{eqnarray}
and has the dimension of an inverse energy. In the limits Eqs.~(\ref{G_limit})---(\ref{G_limit_th}), where the PSN becomes thermal Gaussian noise, the constant $a$ is identical with the inverse temperature $\beta$, so that $f_\tau(p)$ then agrees with the fluctuation function considered e.g. in \cite{VanZon03}.

\begin{figure}
\begin{center}
\includegraphics[width=8cm]{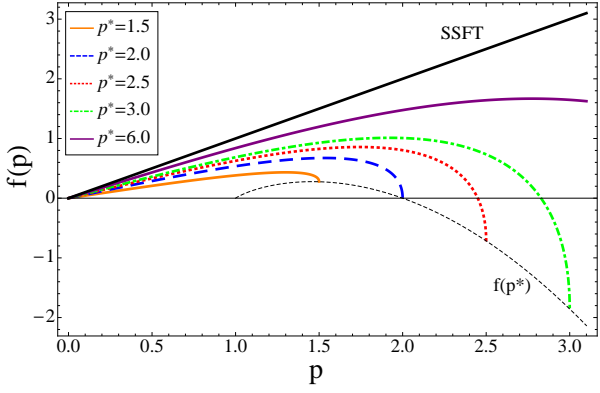}
\caption{\label{Fig_ft1}The fluctuation function $f(p)$, Eq.~(\ref{fluc_func}), plotted as a function of $p$ for various $p^*$ values in regime (i). For $p^*>2$, $f(p)$ becomes negative if $p\in[p_+,p^*]$. The approach to $f(p^*)$ is with a vertical slope, while $f(p^*)$ itself remains finite, Eq.~(\ref{end-points}) (thin dashed curve). For large $p^*$, $f(p)$ approaches the straight line of the SSFT.}
\end{center}
\end{figure}

\begin{figure}
\begin{center}
\includegraphics[width=8cm]{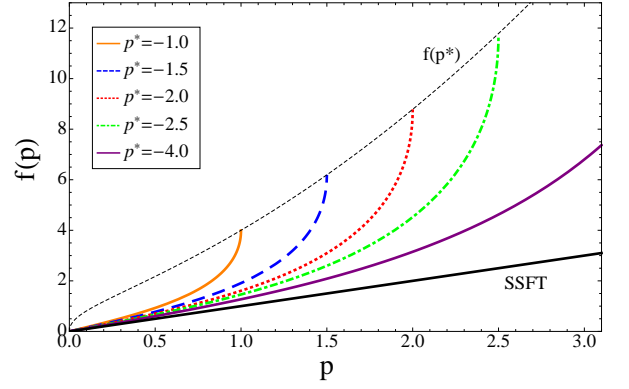}
\caption{\label{Fig_ft2}The fluctuation function $f(p)$, Eq.~(\ref{fluc_func}), plotted as a function of $p$ for various $p^*$ values in regime (ii). Using the antisymmetry of $f(p)$ with respect to $p$ we plot only $p\in[0,-p^*]$. The approach to $f(p^*)$ is with a vertical slope, while $f(p^*)$ itself remains finite, Eq.~(\ref{end-points}) (thin dashed curve). For large $p^*$, $f(p)$ approaches the straight line of the SSFT.} 
\end{center}
\end{figure}

The steady state fluctuation theorem (SSFT), Eq.~(\ref{conventional}), predicts that $f(p)\equiv\lim_{\tau\rightarrow\infty}f_\tau(p)=p$. From Eq.~(\ref{largeDev}) we obtain here instead
\begin{eqnarray}
\label{fluc_func}
f(p)&=&2(p^*-1)p\nonumber\\
&&+2(p^*-1)^2\left(\sqrt{\frac{p^*-p}{p^*-1}}-\sqrt{\frac{p^*+p}{p^*-1}}\right),
\end{eqnarray}
defined on the interval $[-p^*,p^*]$ \cite{Footnote2}. We see that the SSFT is violated in our model, even though we have identified a large deviation form of the distribution. As $p\rightarrow p^*$ the fluctuation function diverges like $|p^*-p|^{-1/2}$, i.e., the cut-off is approached with a vertical slope (cf. Figs.~\ref{Fig_ft1} and~\ref{Fig_ft2}). However, $f(p)$ itself remains finite at the cut-off and assumes the value
\begin{eqnarray}
\label{end-points}
f(p^*)=2(p^*-1)p^*-2(p^*-1)^2\sqrt{\frac{2p^*}{p^*-1}}.
\end{eqnarray}

We now characterize the behavior of the asymptotic fluctuation function $f(p)$, Eq.~(\ref{fluc_func}), for the three different regimes.

(i) $v>0$ (see Fig.~\ref{Fig_ft1}). Here, $f(p)$ of Eq.~(\ref{fluc_func}) has three zeros at $p_0=0$ and $p_\pm=\pm2\sqrt{p^*-1}$ (due to the antisymmetry of $f(p)$ we need not discuss the negative root $p_-$). The zero $p_+$ becomes significant when $p^*>2$, because then $p^*>p_+$ and $p$ can assume values in the interval $[p_+,p^*]$. The crucial observation is then that $f(p)$ from Eq.~(\ref{fluc_func}) becomes negative for $p\in[p_+,p^*]$. There exists therefore a parameter regime in which negative fluctuations of a certain magnitude are \textit{more likely to occur} than corresponding positive ones (see Fig.~\ref{Fig_ft1}). In fact, since $p^*=1+\tau_p/\tau_\lambda$ we find that $p^*>2$ if $\tau_p>\tau_\lambda$.

This property is due to the strongly asymmetric tails of the work distribution $\Pi_\tau(p)$: the negative tail decays exponentially, while the positive tail decays more rapidly to the cut-off at $p^*$, so that $\Pi_\tau(p)<\Pi_\tau(-p)$ for $p$ values in the vicinity of $p^*$ (cf. the discussion of the rate function in regime (i), below Eq.~(\ref{largeDev})). It is important to note that despite the existence of a considerable negative regime of $f(p)$, the second law is never violated: the mean value of the work is always positive, $\left<W_\tau\right>=\alpha v^2\tau$.

(ii) $v<0$ and $\tau_p>\tau_\lambda$ (see Fig.~\ref{Fig_ft2}). In this case $f(p)$ is zero only at $p=0$ and increases monotonically for $p\rightarrow p^*$ until the finite value $f(p^*)$ is reached. For large $p^*$, $f(p)$ approaches the straight line of the SSFT.

(iii) $v<0$ and $\tau_p<\tau_\lambda$. In this parameter regime no negative work fluctuations occur and the fluctuation function $f_\tau(p)$, Eq.~(\ref{fluctuations}) can not be defined.

For both $v<0$ and $v>0$ we observe a pronounced linear regime of $f(p)$ for small $p$ values. This is a general consequence of the large deviation form of $\Pi_\tau(p)$: expanding the rate function $I(p)$ around $p=0$ and taking the ratio $\Pi_\tau(p)/\Pi_\tau(-p)$ leads to a cancelation of the quadratic orders and therefore the linear term dominates up to order $p^3$ in the fluctuation function (cf. \cite{Touchette08}).

\subsection{Gaussian limit and the SSFT}
\label{Sec_Gauss}

In the limits Eqs.~(\ref{G_limit})---(\ref{G_limit_th}) PSN goes over into Gaussian noise. In this limit we should therefore reproduce previous results for a dragged Brownian particle in a parabolic potential (see e.g. \cite{VanZon03,Taniguchi07}). In particular, we should obtain a Gaussian work distribution and the work fluctuations should satisfy the SSFT, Eq.~(\ref{conventional}).

In order to demonstrate this we expand the characteristic function of work, Eq.~(\ref{cf_final}) in powers of $\Gamma_0$. Retaining terms up to second order in $\Gamma_0$ yields
\begin{eqnarray}
\ln G_{W_\tau}(q)&\approx&iq\alpha v^2\tau\nonumber\\
&&-q^2\Gamma^2_0v^2\left(\frac{\tau}{\tau_\lambda}-\frac{\tau_r}{\tau_\lambda}(1-e^{-\tau/\tau_r})\right).
\end{eqnarray}
This means that in the Gaussian limit of PSN the work distribution is Gaussian with mean $\left<W_\tau\right>=\alpha v^2\tau$, Eq.~(\ref{mean_work}), and variance
\begin{eqnarray}
\label{gauss_var}
\left<W^2_\tau\right>-\left<W_\tau\right>^2=2\Gamma^2_0v^2\left(\frac{\tau}{\tau_\lambda}-\frac{\tau_r}{\tau_\lambda}(1-e^{-\tau/\tau_r})\right).
\end{eqnarray}
For thermal Gaussian noise the fluctuation-dissipation theorem requires that $\Gamma_0^2=\tau_\lambda\alpha\beta^{-1}$ (cf. Eq.~(\ref{G_limit_th})). In the asymptotic regime we thus obtain from Eq.~(\ref{gauss_var})
\begin{eqnarray}
\label{asymp_var}
\left(\left<W^2_\tau\right>-\left<W_\tau\right>^2\right)\cong2\beta^{-1}\left<W_\tau\right>,
\end{eqnarray}
i.e., the variance is proportional to the mean. The mean Eq.~(\ref{mean_work}) and the variance Eq.~(\ref{asymp_var}) determine the distribution of the work $W_\tau$. For the distribution of the rescaled work $p=W_\tau/\left<W_\tau\right>$ one then obtains the Gaussian
\begin{eqnarray}
\label{gauss_dist}
\Pi_\tau(p)\cong\sqrt{\frac{\beta\left<W_\tau\right>}{4\pi}}e^{-\beta\left< W_\tau\right>(p-1)^2/4},
\end{eqnarray} 
in the asymptotic regime. The saddle-point approximation Eq.~(\ref{p_dist}) yields the same result, if one expands with Eq.~(\ref{p_cutoff}) in powers of $\tau_\lambda/\tau_p$
\begin{eqnarray}
\sqrt{\frac{p^*-p}{p^*-1}}\approx 1-\sigma(v)\frac{1}{2}\frac{\tau_\lambda}{\tau_p}(p-1),
\end{eqnarray}
since $\tau_\lambda/\tau_p\rightarrow 0$ in the Gaussian limit.
The fluctuation function $f_\tau(p)$, defined by Eqs.~(\ref{fluctuations}) and (\ref{const_a}), then satisfies $\lim_{\tau\rightarrow \infty}f_\tau(p)=p$. We therefore confirm that, in the limit where the PSN goes over into thermal Gaussian noise, the SSFT holds.

\section{Inertial effects for finite times}
\label{Sec_inertia}

For finite $\tau$ inertial effects can be significant if $\tau_m$ is of the order of $\tau_r$ or larger. The eigenvalues $\nu_1$ and $\nu_2$, determined by Eq.~(\ref{eigenvalues}), become complex if $\tau_m>\tau_r/4$. This condition can likewise be expressed in terms of a critical mass $m^*$ determined by \cite{Taniguchi08}
\begin{eqnarray}
m^*\equiv\frac{\alpha^2}{4\kappa},
\end{eqnarray}
so that inertial effects become significant when $m>m^*$. For $m>m^*$ the eigenvalues $\nu_1$ and $\nu_2$ of Eq.~(\ref{eigenvalues}) are complex conjugates and can be written as
\begin{eqnarray}
\label{nu12}
\nu_{1,2}=\mu\pm i \omega,
\end{eqnarray}
where we define
\begin{eqnarray}
\mu\equiv -\frac{1}{2\tau_m},\qquad \omega\equiv \frac{1}{2\tau_m}\sqrt{4\tau_m/\tau_r-1}.
\end{eqnarray}
Complex eigenvalues lead to a time oscillatory behavior of the position coordinate $y(t)$, which oscillates with frequency $\omega$. Time oscillations are also manifest in the work distribution, as observed in \cite{Taniguchi08} for Gaussian noise. In \cite{Taniguchi08} it has been shown that for large $\tau$ the oscillation frequency of the fluctuation function is the same as that for the position coordinate (i.e. $\omega$). Since our model is essentially the same damped oscillator as was investigated in \cite{Taniguchi08}, only driven by a different noise, we expect a similar behavior of the work fluctuations for $m>m^*$.

\begin{figure}
\begin{center}
\includegraphics[width=8cm]{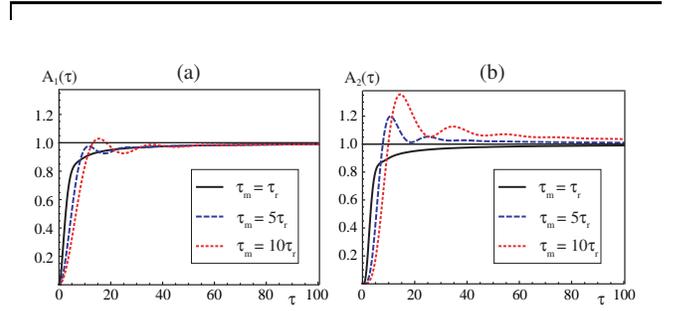}
\caption{\label{Fig_mean}Oscillatory behavior of (a) the rescaled mean $A_1(\tau)$, Eq.~(\ref{mean_resc}), and (b) the rescaled variance $A_2(\tau)$, Eq.~(\ref{var_resc}), for the initial conditions $y_0=u_0=0$ and various $\tau_m$. In the limit $\tau\rightarrow\infty$ both $A_1(\tau)$ and $A_2(\tau)$ converge to $1$. Parameter values: $\tau_r=1$, $\tau_\lambda=0.2$, $\Gamma_0=0.5$.}
\end{center}
\end{figure}

Even though we do not have an analytical expression for the work distribution including inertia, we can investigate inertial effects for the work fluctuations from the characteristic function of the work Eq.~(\ref{cf_work}) via the cumulants of the work distribution $c_n(\tau)$, defined as \cite{VanKampen}
\begin{eqnarray}
\ln G_{W_\tau}(q)=\sum_{n=1}^\infty \frac{(iq)^n}{n!}c_n(\tau).
\end{eqnarray}
We thus obtain the cumulants by calculating the derivatives of $\ln G_{W_\tau}(q)$
\begin{eqnarray}
c_n(\tau)=\left.\frac{1}{i^n}\frac{\partial^n}{\partial q^n}\ln G_{W_\tau}(q)\right|_{q=0},
\end{eqnarray}
which yields from Eq.~(\ref{cf_work}) expressions for $c_1(\tau)$, the mean, and $c_2(\tau)$, the variance of the work distribution:
\begin{widetext}
\begin{eqnarray}
\label{in_mean}
\left<W_\tau\right>&=&\frac{v\,\kappa}{\nu_1-\nu_2}\left[y_0\left(\frac{\nu_1}{\nu_2}\left(1-e^{\nu_2\tau}\right)-\frac{\nu_2}{\nu_1}\left(1-e^{\nu_1\tau}\right)\right)+u_0\left(\frac{1}{\nu_1}\left(1-e^{\nu_1\tau}\right)-\frac{1}{\nu_2}\left(1-e^{\nu_2\tau}\right)\right)\right.\nonumber\\
&&\left.-\frac{\alpha v}{m}
\left(\frac{1}{\nu_1^2}\left(\nu_1\tau+1-e^{\nu_1\tau}\right)-\frac{1}{\nu_2^2}\left(\nu_2\tau+1-e^{\nu_2\tau}\right)\right)\right],\\
\label{in_var}
\left<W^2_\tau\right>-\left<W_\tau\right>^2&=&2\frac{\Gamma_0^2}{\tau_\lambda}\left(\frac{v\,\kappa}{m(\nu_1-\nu_2)}\right)^2
\int_0^\tau\left(\frac{1}{\nu_1}\left(1-e^{\nu_1(\tau-t)}\right)-\frac{1}{\nu_2}\left(1-e^{\nu_2(\tau-t)}\right)\right)^2\upd t,
\end{eqnarray}
and the $n$th-order cumulant reads
\begin{eqnarray}
\label{in_higher}
c_n(\tau)=n!\,\frac{1}{\tau_\lambda}\left(\frac{\Gamma_0 v\,\kappa}{m(\nu_1-\nu_2)}\right)^n
\int_0^\tau\left(\frac{1}{\nu_1}\left(1-e^{\nu_1(\tau-t)}\right)-\frac{1}{\nu_2}\left(1-e^{\nu_2(\tau-t)}\right)\right)^n\upd t,
\end{eqnarray}
respectively. We note that in Eq.~(\ref{in_mean}) the initial position $y_0$ and initial velocity $u_0$ of the particle appear explicitly. We now focus on the behavior of the mean and the variance. Using Eq.~(\ref{nu12}) allows us to rewrite the mean work Eq.~(\ref{in_mean}) in terms of trigonometric functions
\begin{eqnarray}
\label{p_mean}
\left<W_\tau\right>&=&\frac{v\,\kappa}{\mu^2+\omega^2}\left[y_0\left(2\mu+e^{\mu\tau}\left(\frac{\mu^2-\omega^2}{\omega}\sin(\omega\tau)-2\mu\cos(\omega\tau)\right)\right)-u_0\left(1+e^{\mu\tau}\left(\frac{\mu}{\omega}\sin(\omega\tau)-\cos(\omega\tau)\right)\right)\right]\nonumber\\
&&+\frac{\alpha v^2}{\tau_m\tau_r(\mu^2+\omega^2)^2}\left(\tau(\mu^2+\omega^2)+2\mu+e^{\mu\tau}\left(\frac{\mu^2-\omega^2}{\omega}\sin(\omega\tau)-2\mu\cos(\omega\tau)\right)\right).
\end{eqnarray}
Likewise, one can express the variance, Eq.~(\ref{in_var}), in the form
\begin{eqnarray}
\label{p_var}
\left<W^2_\tau\right>-\left<W_\tau\right>^2&=&\frac{2\Gamma_0^2}{\tau_\lambda(\mu^2+\omega^2)^3}\left(\frac{v}{2\tau_m\tau_r\omega}\right)^2\left[4(\omega^2\tau+\mu)(\mu^2+\omega^2)-3\mu(\mu^2-3\omega^2)-\frac{1}{\mu}(\mu^2+\omega^2)^2(1-e^{2\mu\tau})\right.\nonumber\\
&&\left.+ 8\omega e^{\mu\tau}\left((\mu^2-\omega^2)\sin(\omega\tau)-2\mu\omega\cos(\omega\tau)\right)\right.\nonumber\\
&&\left.-e^{2\mu\tau}\left(\mu(\mu^2-3\omega^2)\cos(2\omega\tau)+\omega(3\mu^2-\omega^2)\sin(2\omega\tau)\right)\right].
\end{eqnarray}
\end{widetext}
Both mean and variance show an oscillatory decaying behavior. The mean oscillates with frequency $\omega$ while the variance exhibits oscillations with a superposition of frequencies $\omega$ and $2\omega$. For very large $\tau$ we obtain from Eqs.~(\ref{p_mean}) and (\ref{p_var}):
\begin{eqnarray}
\label{mean_OD}
\left<W_\tau\right>&\cong&\alpha v^2\tau,\\
\label{var_OD}
\left<W^2_\tau\right>-\left<W_\tau\right>^2&\cong&2\Gamma_0^2v^2\tau/\tau_\lambda,
\end{eqnarray}
i.e., we recover the mean and the variance of the overdamped work distribution (cf. Eq.~(\ref{mean_work}) and Eq.~(\ref{gauss_var})) respectively, as expected. The oscillatory behavior of the mean and the variance are shown in Fig.~\ref{Fig_mean}, where we plot the rescaled quantities
\begin{eqnarray}
\label{mean_resc}
A_1(\tau)&\equiv& \frac{\left<W_\tau\right>}{\alpha v^2\tau},\\
\label{var_resc}
A_2(\tau)&\equiv&\frac{\left<W^2_\tau\right>-\left<W_\tau\right>^2}{2\Gamma_0^2v^2\tau/\tau_\lambda},
\end{eqnarray}
using Eqs.~(\ref{mean_OD}) and~(\ref{var_OD}), respectively. Both $A_1$ and $A_2$ converge to $1$ in the limit $\tau\rightarrow\infty$.

Considering the higher order cumulants, one can also rewrite the $n$th-order cumulant Eq.~(\ref{in_higher}) in terms of trigonometric functions. Without performing the calculation, one notices that this would lead to decaying oscillations with a superposition of frequencies $\omega,2\omega,3\omega,...,n\omega$.

\section{Superposition of PSN and thermal Gaussian noise}
\label{Sec_GP}

So far we have investigated the properties of the work fluctuation of the dragged particle of Sec.~\ref{Sec_model}, when the noise from the environment is given purely by external PSN. In this section we investigate the effect of additional thermal Gaussian noise on the fluctuation properties of the particle. The additional Gaussian noise takes into account the effect of an additional equilibrium heat bath on the dynamics. This model can therefore represent a dragged \textit{Brownian particle}, which is subject to PSN. For a Brownian particle the thermal fluctuations and the friction have the same physical origin, namely the surrounding heat bath (`water'), so that the friction and the noise strength of the thermal noise are related via a fluctuation-dissipation relation. On the other hand, we assume that the PSN arises due to an external physical mechanism that is independent of the thermal noise, which implies that the two types of noises are statistically independent. 

Gaussian noise is symmetric so that the particle will now be able to access all positions in the harmonic potential. Both the position and the work cut-off of the purely PSN case, discussed in Sec.~\ref{Sec_sym}, are therefore expected to disappear. Our quantitative investigations start from the equation of motion for the comoving coordinate $y$ in the overdamped regime
\begin{eqnarray}
\label{y_gp}
\dot{y}(t)=-\frac{1}{\tau_r}y(t)-v_e+\frac{1}{\alpha}z(t)+\frac{1}{\alpha}\eta(t),
\end{eqnarray}
which is Eq.~(\ref{F_y}) with additional thermal Gaussian noise $\eta(t)$
\begin{eqnarray}
\left<\eta(t)\right>&=&0,\\
\label{Gauss_noise}
\left<\eta(t)\eta(t')\right>&=&2\alpha\beta^{-1}\delta(t-t'),
\end{eqnarray}
where $\beta$ is interpreted as inverse temperature of the equilibrium heat bath. Eq.~(\ref{Gauss_noise}) expresses the fluctuation-dissipation relation between friction and noise strength \textbf{\cite{Footnote4}}. The characteristic noise functional of $\eta(t)$ is given by \cite{VanKampen}
\begin{eqnarray}
\label{G_nf}
G_{\eta(t)}\left[g(t)\right]&=&\exp\left\{-\frac{\alpha}{\beta}\int_0^\infty g(t)^2\upd t\right\},
\end{eqnarray}
for a test function $g(t)$. The characteristic functional of $y(t)$ can then be calculated analogous to the two-component case treated in Sec.~\ref{Sec_cf}, using the theorem of C\'aceres-Budini \cite{Caceres97}. One obtains (cf. Eq.~(\ref{cb})) \cite{Baule}
\begin{eqnarray}
\label{cb_pg}
G_{y(t)}[h(t)]&=&e^{iy_0k_0-iv_e\int_0^\infty k(t)\upd t}G_{z(t)}[k(t)/\alpha]\nonumber\\
&&\times G_{\eta(t)}[k(t)/\alpha],
\end{eqnarray}
where $k(t)$ is given by
\begin{eqnarray}
\label{k_pg}
k(t)=\int_t^\infty e^{(t-s)/\tau_r} h(s)\upd s,
\end{eqnarray}
and $y_0=y(t=0)$ as well as $k_0=k(t=0)$. The superposition of the two statistically independent noises $z(t)$ and $\eta(t)$ in Eq.~(\ref{y_gp}) leads therefore to the product of the corresponding noise functionals in Eq.~(\ref{cb_pg}). Substituting the noise functionals $G_{z(t)}$, Eq.~(\ref{noise_functional}), and $G_{\eta(t)}$, Eq.~(\ref{G_nf}), with the test function $k(t)/\alpha$ as argument into Eq.~(\ref{cb_pg}) yields
\begin{widetext}
\begin{eqnarray}
\label{y_cf_pg2}
G_{y(t)}[h(t)]=\exp\left\{iy_0k_0-iv_e\int_0^\infty k(t)\upd t-\frac{1}{\alpha\beta}\int_0^\infty k(t)^2\upd t+\frac{1}{\tau_\lambda}\int_0^\infty\left(\frac{1}{1-i\frac{\Gamma_0}{\alpha}k(t)}-1\right)\upd t\right\}.
\end{eqnarray}
\end{widetext}
From this characteristic functional with $k(t)$ given by Eq.~(\ref{k_pg}) we can determine both the characteristic function of the particle position and that of the work by choosing appropriate test functions $h(t)$.

\subsection{Distribution of the particle position in the NESS}

\begin{figure}
\begin{center}
\includegraphics[width=8cm]{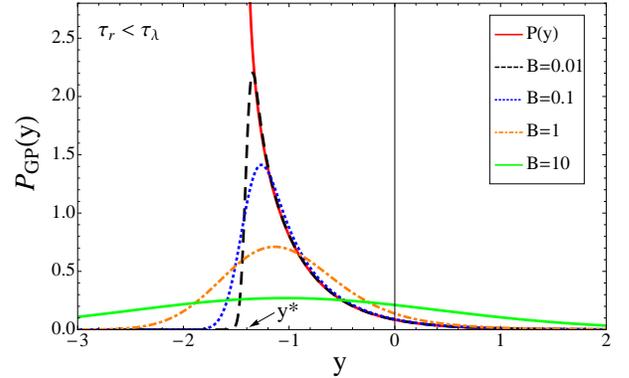}
\caption{\label{Fig_dist_ygp2}The distributions of the particle position $P_{GP}(y)$, Eq.~(\ref{conv}), and $P(y)$, Eq.~(\ref{p_stat}) for $\tau_r<\tau_\lambda$ and various values of the noise ratio $B$. $P(y)$ exhibits a divergence at the cut-off $y^*=-1.4$, which is reduced to a maximum in $P_{GP}(y)$ due to the additional Gaussian noise. For $B>0$ the left tails of the various curves are Gaussian and the right tails exponential. Parameter values: $\tau_r=1$, $\tau_\lambda=1.25$, $\Gamma_0=0.5$, $v=1$, $\tau_r=1$. }
\end{center}
\end{figure}

\begin{figure}
\begin{center}
\includegraphics[width=8cm]{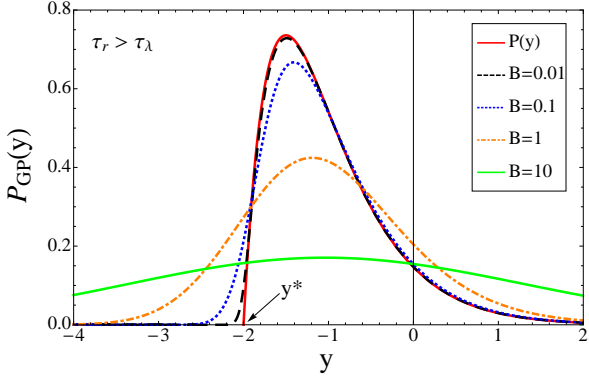}
\caption{\label{Fig_dist_ygp1}The distributions of the particle position $P_{GP}(y)$, Eq.~(\ref{conv}), and $P(y)$, Eq.~(\ref{p_stat}), for $\tau_r>\tau_\lambda$ and various values of the noise ratio $B$. For small $B$ values $P_{GP}(y)$ approaches $P(y)$ (solid red curve), which exhibits a cut-off at $y^*=-2.0$. For $B>0$ the left tails of the various curves are Gaussian and the right tails exponential. Parameter values: $\tau_r=1$, $\tau_\lambda=0.5$, $\Gamma_0=0.5$, $v=1$.}
\end{center}
\end{figure}

The characteristic function of the particle position is obtained from $G_{y(t)}[h(t)]$ if we choose the test function $h(t)= h_1\delta(t-t_1)$ (cf. Appendix~\ref{App_gp}). The distribution of the particle position in the NESS is then obtained by carrying out an inverse Fourier-transform of the characteristic function. Details of this calculation are presented in Appendix~\ref{App_gp}. The result for the distribution can be written in the form of the convolution integral
\begin{eqnarray}
\label{conv}
P_{GP}(y)&=& \int_{-\infty}^\infty P(y')P_G(y-y')\upd y',
\end{eqnarray}
where $P(y)$ and $P_G(y)$ are the NESS distributions of the particle position in the purely PSN case and purely Gaussian case, respectively. $P(y)$ is given by Eq.~(\ref{p_stat}) and $P_G(y)$ is given by the Gaussian
\begin{eqnarray}
P_G(y)=\sqrt{\frac{\beta\kappa}{2\pi}}e^{-\frac{\beta\kappa}{2}y^2}.
\end{eqnarray}
The superposition of the two independent noises in the Langevin equation~(\ref{y_gp}) thus gives rise to a convolution of the corresponding distributions. A closed form expression of the integral Eq.~(\ref{conv}) is given in Appendix~\ref{App_gp} in Eq.~(\ref{pg_dist_closed}).

In order to characterize the relative strength of the PSN to the Gaussian noise we introduce the dimensionless noise ratio $B$, defined as
\begin{eqnarray}
B\equiv\frac{\alpha\tau_\lambda}{\beta\Gamma_0^2},
\end{eqnarray}
which represents the ratio of the noise strength of the Gaussian noise, $2\alpha/\beta$ (Eq.~(\ref{Gauss_noise})), and the noise strength of the PSN $z(t)$, $2\Gamma_0^2/\tau_\lambda$ (cf. Eq.~(\ref{PSN_noise2})). This means that large $B$ indicates a dominant influence of the Gaussian noise and small $B$ that of PSN. The limit case $B=0$ corresponds to the purely PSN case (see below).

We plot $P_{GP}(y)$ separately for $\tau_r<\tau_\lambda$ and $\tau_r>\tau_\lambda$, in the Figs.~\ref{Fig_dist_ygp2} and \ref{Fig_dist_ygp1} respectively, for different values of $B$.

In the case $\tau_r<\tau_\lambda$ (see Fig.~\ref{Fig_dist_ygp2}) the distribution $P(y)$ of Eq.~(\ref{p_stat}) exhibits a divergence at the position cut-off $y^*$ (cf. solid red curve in Fig.~\ref{Fig_dist_ygp2}). Under the superimposed Gaussian noise this divergence reduces to a maximum which shifts more and more to $y^*$ the smaller the values of $B$. At the same time one notices that in the approach to the maximum the curves of $P_{GP}(y)$ for $B=0.1$ and $B=0.01$ are basically on top of $P(y)$ (cf. dashed black and dotted blue curves in Fig.~\ref{Fig_dist_ygp2}), i.e., the right tail of $P^{GP}(y)$ approaches $P(y)$ for small $B$. The left tail on the other hand decays like a Gaussian

In the case $\tau_r>\tau_\lambda$ (see Fig.~\ref{Fig_dist_ygp1}) there is no divergence in the distribution $P(y)$. For small $B$, $P_{GP}(y)$ approaches the shape of $P(y)$, yet without exhibiting a cut-off. This can be seen in the curve of $P_{GP}(y)$ for $B=0.01$ (cf. dashed black curve in Fig.~\ref{Fig_dist_ygp1}), which lies on top of the curve of $P(y)$ (solid red curve) apart from a region in the vicinity of $y^*$. For any $B>0$ the left tail of $P_{GP}(y)$ always extends beyond $y^*$ and decays like a Gaussian, indicating that the position cut-off vanishes due to the additional Gaussian noise.

In both cases $P_{GP}(y)$ becomes broader and broader for increasing $B$, i.e., stronger Gaussian noise (higher temperature).

\subsection{Work fluctuations}

From the characteristic functional $G_{y(t)}[h(t)]$, Eq.~(\ref{y_cf_pg2}), we obtain the characteristic function of the work $G_{W_\tau}(q)$ in a similar way as in the two-component case treated in Sec.~\ref{Sec_cf}, by considering the test function (cf. Appendix~\ref{App_sp})
\begin{eqnarray}
\label{tilde_h}
\tilde{h}(t)=- q v \kappa\Theta(\tau-t).
\end{eqnarray}
Substituting $\tilde{h}(t)$ for $h(t)$ into Eq.~(\ref{y_cf_pg2}) leads to the characteristic function of the work
\begin{eqnarray}
\label{gp_cf}
G_{W_\tau}(q)&=&\left(1+iq\Gamma_0 v\right)^\frac{\tau_r/\tau_\lambda}{1+iq \Gamma_0 v }\exp\left\{iqW_\tau^*
\left(1-\frac{\tau_r}{\tau}\right)\right.\nonumber\\
&&\left.-q^2\frac{\left<W\right>}{\beta}\left(1-\frac{3}{2}\frac{\tau_r}{\tau}\right)\right.\nonumber\\
&&\left.+\frac{\tau}{\tau_\lambda}\left(\frac{1}{1+iq \Gamma_0 v }-1\right)\right\},
\end{eqnarray}
upon neglecting exponential terms in $\tau$ and choosing the particular initial condition $y_0=0$. In contrast to the purely PSN case, where the initial position $y_0$ has been sampled from the NESS distribution (cf. Sec.~\ref{Sec_cf}), we consider here a fixed initial position for simplicity. In the $\tau\rightarrow \infty$ limit, in which we are interested in here, the particular initial condition is irrelevant for the properties of the work fluctuations, because the work $W_\tau$ is extensive in $\tau$ (cf. Eq.~(\ref{work_par})). Therefore, a fixed initial condition yields the same result for the rate function as an average over the initial states. 

The distribution of the rescaled work $p$ is then given by the inverse Fourier transform of $G_{W_\tau}(q)$. This Fourier inversion can be performed as in Sec.~\ref{Sec_FT}, using the method of steepest descent. Details of this calculation are presented in Appendix~\ref{App_sp}. The saddle-point $\bar{q}$ is now given by
\begin{eqnarray}
\label{gp_q}
\bar{q}=\frac{i}{\Gamma_0 v}r_p,
\end{eqnarray}
where $r_p$ is determined by solving the cubic equation
\begin{eqnarray}
\label{cubic}
2B\, r_p(r_p-1)^2-\left(\frac{p^*-p}{p^*-1}\right)(r_p-1)^2+1=0,
\end{eqnarray}
where, $p^*$ denotes the work cut-off of the purely PSN model, Eq.~(\ref{p_cutoff}). Since there is no restriction on the possible $p$ values in the presence of the additional Gaussian noise, $p^*$ does not have the meaning of a cut-off here, but nevertheless appears as a parameter characterizing the PSN. An analysis of Eq.~(\ref{cubic}) shows that this cubic equation has a unique real root $<1$ for all possible values of $B$, $p^*$, and $p$ (see Appendix \ref{App_root}). This particular root (denoted by $r_p$ in the following) thus yields the correct saddle-point $\bar{q}$, since the integration path in the saddle-point approximation can be deformed to go through $\bar{q}$ without crossing the pole at $q=i/(\Gamma_0v)$ in the characteristic function Eq.~(\ref{gp_cf}).

Using $\bar{q}$ of Eq.~(\ref{gp_q}) with the particular root $r_p$, the saddle-point approximation of the work distribution $\Pi^{GP}_\tau(p)$ can be obtained in a straightforward way and reads (cf. Appendix~\ref{App_sp})
\begin{widetext}
\begin{eqnarray}
\label{gp_work_dist}
\Pi^{GP}_\tau(p)&\cong&\sqrt{\frac{\tau/\tau_\lambda}{4\pi (1+B(1-r_p)^3)}}\frac{(1-r_p)^{\frac{\tau_r/\tau_\lambda}{1-r_p}+\frac{3}{2}}}{|p^*-1|}\nonumber\\
&&\times\exp\left\{-\frac{\tau}{\tau_\lambda}\left[r_p\left(\frac{p^*-p}{p^*-1}\right)\left(1-\frac{\tau_r}{\tau}\right)-r_p^2B\left(1-\frac{3\tau_r}{2\tau}\right)-\left(\frac{1}{1-r_p}-1\right)\right]\right\}.
\end{eqnarray}
\end{widetext}
Noting that $p^*=1+\sigma(v)\tau_p/\tau_\lambda$, Eq.~(\ref{p_cutoff}), $\Pi^{GP}_\tau(p)$ is completely specified by the times $\tau,\tau_r,\tau_\lambda,\tau_p$ and the parameter $B$. Furthermore, since $r_p<1$ always, the work distribution is real for all $p$ values, so that there is no work cut-off as expected.

We now investigate the behavior of the work distribution in more detail in the asymptotic regime $\tau\rightarrow\infty$, where $\Pi^{GP}_\tau(p)$ assumes the large deviation form
\begin{eqnarray}
\label{gp_ldev}
\Pi^{GP}_\tau(p)\cong e^{-\tau I^{GP}(p)}
\end{eqnarray}
with rate function
\begin{eqnarray}
\label{rf_gp}
I^{GP}(p)=\frac{1}{\tau_\lambda}\left[r_p\left(\frac{p^*-p}{p^*-1}\right)-r_p^2B-\left(\frac{1}{1-r_p}-1\right)\right].
\end{eqnarray}

\begin{figure}
\begin{center}
\includegraphics[width=7cm]{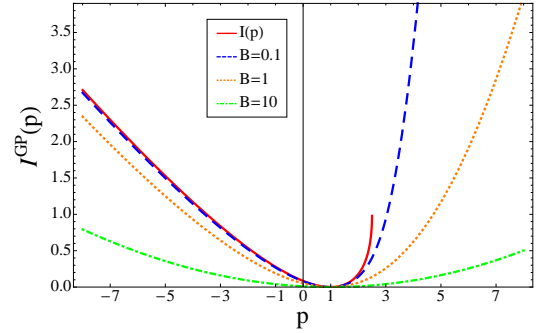}
\caption{\label{Fig_rf_gp1}The rate function $I^{GP}(p)$, Eq.~(\ref{rf_gp}), plotted as a function of $p$ for various $B$ values together with $I(p)$, Eq.~(\ref{rate_func}), in regime (i). The right tail of $I(p)$ ends at the cut-off value $p^*=2.5$, where $I(p^*)=1.0$, while that of $I^{GP}(p)$ increases monotonically for any $B>0$. Asymptotically the left tail of $I^{GP}(p)$ becomes linear and the right tail quadratic. Parameter values: $\tau_\lambda=1.0$.}
\end{center}
\end{figure}

\begin{figure}
\begin{center}
\includegraphics[width=7cm]{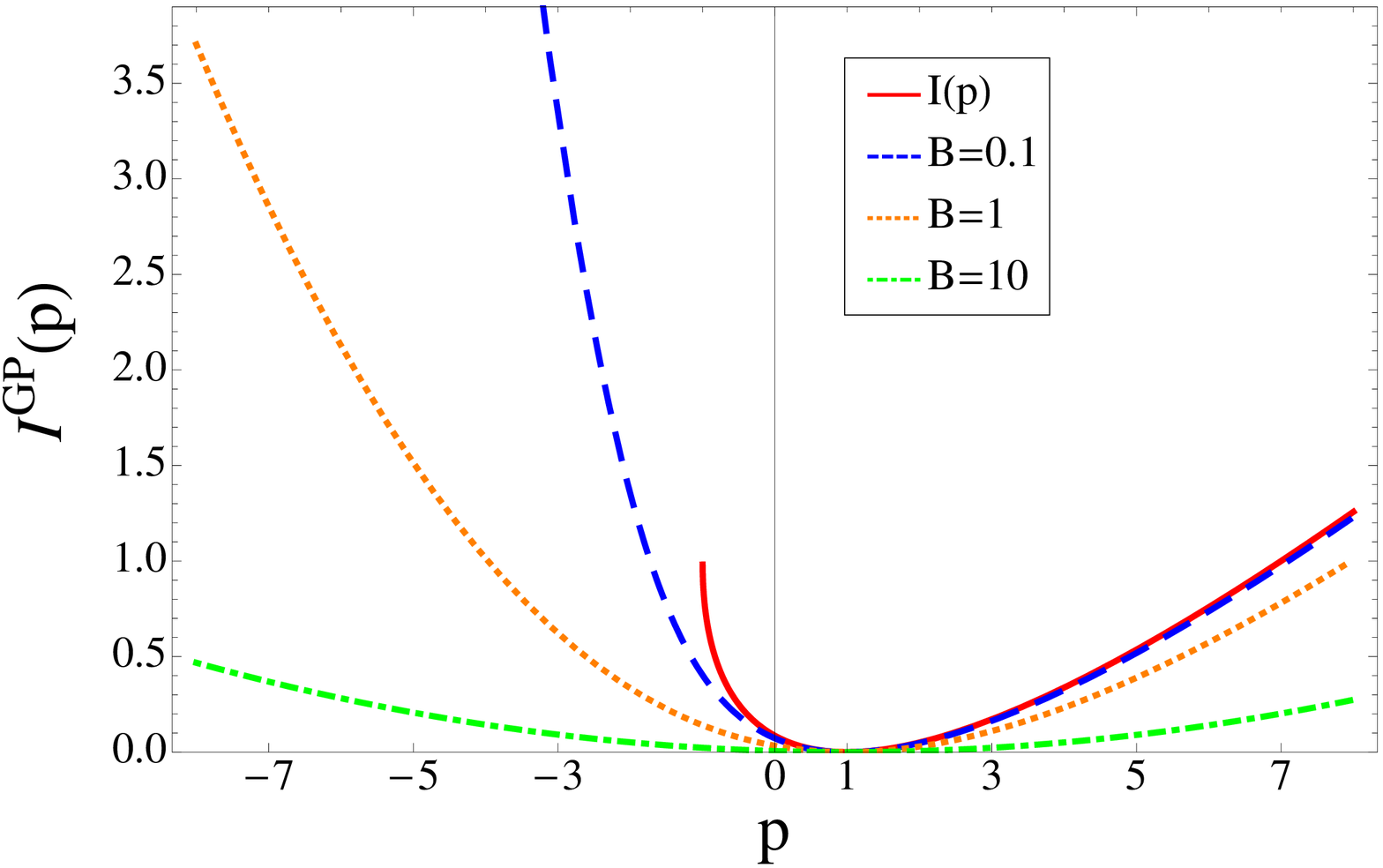}
\caption{\label{Fig_rf_gp2}The rate function $I^{GP}(p)$, Eq.~(\ref{rf_gp}), plotted as a function of $p$ for various $B$ values together with $I(p)$, Eq.~(\ref{rate_func}), in regime (ii). The left tail of $I(p)$ ends at the cut-off value $p^*=-1.0$, where $I(p^*)=1.0$, while the left tail of $I^{GP}(p)$ increases monotonically for any $B>0$. Asymptotically the left tail of $I^{GP}(p)$ becomes quadratic and the right tail linear. Parameter values: $\tau_\lambda=1.0$.}
\end{center}
\end{figure}

\begin{figure}
\begin{center}
\includegraphics[width=7cm]{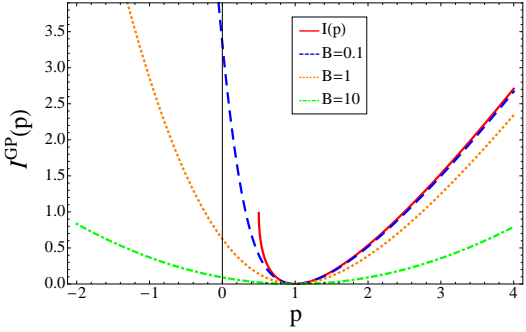}
\caption{\label{Fig_rf_gp3}The rate function $I^{GP}(p)$, Eq.~(\ref{rf_gp}), plotted as a function of $p$ for various $B$ values together with $I(p)$, Eq.~(\ref{rate_func}), in regime (iii). The left tail of $I(p)$ ends at the cut-off value $p^*=0.5$, where $I(p^*)=1.0$, so that no negative work fluctuations occur in the purely PSN case. On the other hand, the left tail of $I^{GP}(p)$ increases monotonically for any $B>0$ and becomes asymptotically quadratic, while the right tail becomes asymptotically linear. Parameter values: $\tau_\lambda=1.0$.}
\end{center}
\end{figure}

In Figs.~\ref{Fig_rf_gp1}---\ref{Fig_rf_gp3} we plot $I^{GP}(p)$ for various $B$ values together with the rate function $I(p)$ of the purely PSN case, Eq.~(\ref{rate_func}). As before we distinguish three different regimes of the work fluctuations, namely (i) $v>0$ (Fig.~\ref{Fig_rf_gp1}); (ii) $v<0$ and $\tau_p>\tau_\lambda$ (Fig.~\ref{Fig_rf_gp2}); (iii) $v<0$ and $\tau_p<\tau_\lambda$ (Fig.~\ref{Fig_rf_gp3}). In all three cases $I^{GP}(p)$ is asymmetric around its minimum at $p=1$ and becomes broader for increasing $B$. One observes that $I^{GP}(p)$ always extends beyond the work cut-off of the purely PSN case. This becomes particularly evident when comparing the $B=0.1$ curve of $I^{GP}(p)$ (dashed blue curve in Figs.~\ref{Fig_rf_gp1}---\ref{Fig_rf_gp3}) with $I(p)$ (solid red curve). $I^{GP}(p)$ lies virtually on top of $I(p)$ on the unbounded side of $I(p)$, but clearly deviates on the bounded side of $I(p)$ and eventually increases monotonically beyond the cut-off. Moreover, in case (iii) no negative work fluctuations can occur in the purely PSN case due to the positive minimum work cut-off (cf. Eq.~(\ref{p_cutoff})), but $I^{GP}(p)$ does indeed assume values for negative $p$. This means that in case (iii) negative work fluctuations arise due to the additional Gaussian noise.

The behavior in the tails of $I^{GP}(p)$ for $p\rightarrow\pm\infty$ can be determined from the properties of the saddle-point $\bar{q}$, Eq.~(\ref{gp_q}). The saddle-point approximation namely implies that the slope of the rate function $I^{GP}(p)$ is proportional to the saddle point itself (cf. \cite{Touchette08}), or more precisely (cf. Eq.~(\ref{app_slope}) in Appendix~\ref{App_sp}) 
\begin{eqnarray}
\label{gp_rate_slope}
\frac{\upd}{\upd p} I^{GP}(p)=i\bar{q}\alpha v^2=-\sigma(v)\frac{r_p}{\tau_p},
\end{eqnarray}
using Eqs.~(\ref{gp_q}) and (\ref{tau_p}). With Eq.~(\ref{gp_rate_slope}) one can determine the properties of the rate function as $p\rightarrow\pm\infty$ as follows.

First, we consider the case $v>0$. As discussed in Appendix \ref{App_root}, $r_p$ has the properties that $r_p\rightarrow 1$ for $p\rightarrow-\infty$ and $r_p\propto -p$ for $p\rightarrow\infty$ if $v>0$. It then follows from Eq.~(\ref{gp_rate_slope}) that $I^{GP}(p)$ has an asymptotically linear left tail and an asymptotically quadratic right tail. Therefore, from this behavior of the rate function one can conclude that the left tail of the work distribution $\Pi^{GP}_\tau(p)$ for large $\tau$ is asymptotically exponential, while the right tail is asymptotically Gaussian, if $v>0$.

For $v<0$ the situation is reversed because here $r_p\propto p$ for $p\rightarrow-\infty$ and $r_p\rightarrow 1$ for $p\rightarrow\infty$ (see Appendix \ref{App_root}). The left tail of $I^{GP}(p)$ is thus asymptotically quadratic and the right tail asymptotically linear. Consequently, the left tail of $\Pi^{GP}_\tau(p)$ for large $\tau$ is asymptotically Gaussian, while the right tail is asymptotically exponential, if $v<0$.

One also notices in Figs.~\ref{Fig_rf_gp1}---\ref{Fig_rf_gp3} that the rate function $I(p)$ of the purely PSN case represents a discontinuous limit of the combined PSN and Gaussian case. For any arbitrarily small non-zero $B$, both left and right tails of $I^{GP}(p)$ increase monotonically, while for $B=0$, i.e., when $I^{GP}(p)=I(p)$, the rate function ends at a finite point for $p=p^*$. This singular limit is due to the cut-off singularity in the purely PSN case.

In order to further characterize the fluctuation properties we define the dimensionless fluctuation function
\begin{eqnarray}
\label{gp_ff_def}
f^{GP}_\tau(p)\equiv\frac{1}{b\left<W_\tau\right>}\ln\frac{\Pi^{GP}_\tau(p)}{\Pi^{GP}_\tau(-p)},
\end{eqnarray}
where the constant $b$ is defined as
\begin{eqnarray}
b\equiv\frac{\beta}{1+\frac{1}{B}}.
\end{eqnarray}
The reasoning behind this definition is that $b\rightarrow a$ (Eq.~(\ref{const_a})) for $B\rightarrow 0$ and that $b\rightarrow \beta$ for $B\rightarrow \infty$, i.e., we recover the fluctuation functions of the purely PSN case and the Gaussian case respectively, in the corresponding limits of the noise ratio $B$. In the asymptotic time regime the fluctuation function $f^{GP}(p)\equiv \lim_{\tau\rightarrow\infty}f^{GP}_\tau(p)$ is given as
\begin{eqnarray}
\label{ff_gp}
f^{GP}(p)&=&(p^*-1)^2(1+B)\left[\frac{p}{p^*-1}(r_{-p}+r_p)\right.\nonumber\\
&&\left.+\frac{p^*}{p^*-1}(r_{-p}-r_p)+(r_p^2-r_{-p}^2)B\right.\nonumber\\
&&\left.+\frac{1}{1-r_p}-\frac{1}{1-r_{-p}}\right],
\end{eqnarray}
which is obtained by substituting the large deviation form Eq.~(\ref{gp_ldev}) with the rate function Eq.~(\ref{rf_gp}) into Eq.~(\ref{gp_ff_def}). The fluctuation function Eq.~(\ref{ff_gp}) is a function of $p$, $p^*$, and $B$ only. We discuss the behavior of $f^{GP}(p)$ separately for the three different regimes mentioned above.

\begin{figure}
\begin{center}
\includegraphics[width=7cm]{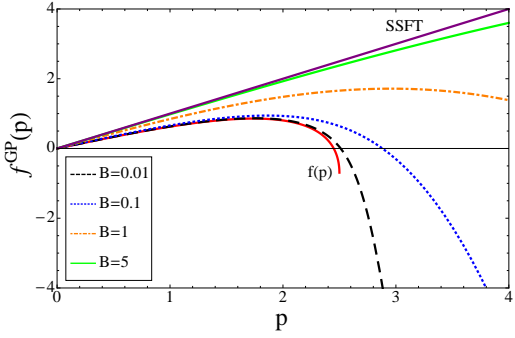} 
\caption{\label{Fig_ft_gp1}The asymptotic fluctuation function $f^{GP}(p)$, Eq.~(\ref{ff_gp}), plotted as a function of $p$ for various $B$ values together with $f(p)$, Eq.~(\ref{fluc_func}), in regime (i). The different behavior in the tails of $\Pi^{GP}_\tau(p)$ leads to a pronounced negative regime of $f^{GP}(p)$, which decreases monotonically for large $p$, while $f(p)$ (solid red curve) ends at a finite value when the cut-off $p^*=2.5$ is reached. For large $B$ the SSFT is recovered (solid purple curve).}
\end{center}
\end{figure}

(i) $v>0$ (Fig.~\ref{Fig_ft_gp1}). For large $B$, i.e., strong thermal Gaussian noise relative to the PSN, $f^{GP}(p)$ becomes linear with slope $1$ in agreement with the SSFT. For small $B$ values, i.e., when the Gaussian noise is weak, $f^{GP}(p)$ approaches the fluctuation function of the purely PSN case, $f(p)$ of Eq.~(\ref{fluc_func}). The important observation is that $f^{GP}(p)$ is negative for $p$ greater than a certain $p_0$, where $p_0$ denotes the value at which $f^{GP}(p)$ intersects the $p$-axis. However, in contrast to the purely PSN case where the fluctuation function $f(p)$ is bounded by $f(p^*)$ (cf. Fig.~\ref{Fig_ft1}), here $f^{GP}(p)\rightarrow-\infty$ as $p\rightarrow\infty$. This behavior is due to the tails of the work distribution $\Pi^{GP}_\tau(p)$ in the asymptotic regime as discussed above: the right tail of $\Pi^{GP}_\tau(p)$ is Gaussian and thus decays more rapidly than the exponential left tail. Consequently $\Pi^{GP}_\tau(p)$ becomes increasingly smaller than $\Pi^{GP}_\tau(-p)$ for increasing $p$, so that $f^{GP}(p)$ is monotonically decreasing for large $p$. The negative regime of the fluctuation function is thus even more pronounced than in the purely PSN case (cf. solid red curve in Fig.~\ref{Fig_ft_gp1}), where the possible work values are bounded by the cut-off $p^*$.

\begin{figure}
\begin{center}
\includegraphics[width=7cm]{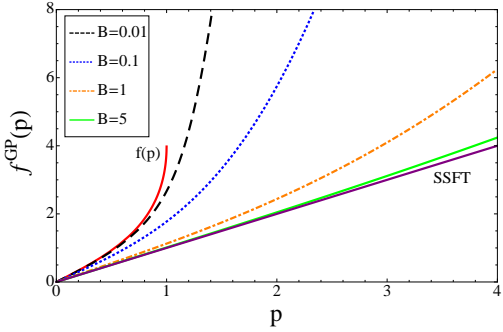} 
\caption{\label{Fig_ft_gp2}The asymptotic fluctuation function $f^{GP}(p)$, Eq.~(\ref{ff_gp}), plotted as a function of $p$ for various $B$ values together with $f(p)$, Eq.~(\ref{fluc_func}), in regime (ii). $f^{GP}_\tau(p)$ increases monotonically for large $p$, while $f(p)$ (solid red curve) ends at a finite value when the cut-off $p^*=1.0$ is reached. For large $B$ the SSFT is recovered (solid purple curve).}
\end{center}
\end{figure}

(ii) $v<0$ with $\tau_p>\tau_\lambda$ (Fig.~\ref{Fig_ft_gp2}). Here, $f^{GP}(p)$ is always positive for $p>0$. For large $B$ values one recovers the SSFT as expected. For small $B$ values $f^{GP}(p)$ approaches the fluctuation function of the purely PSN case. For any $B>0$, $f^{GP}(p)$ is monotonically increasing due to the different behavior in the tails of the work distribution $\Pi^{GP}_\tau(p)$, which behave now oppositely to regime (i): the left tail of $\Pi^{GP}_\tau(p)$ is now Gaussian and decays faster than the exponential right tail.

\begin{figure}
\begin{center}
\includegraphics[width=7cm]{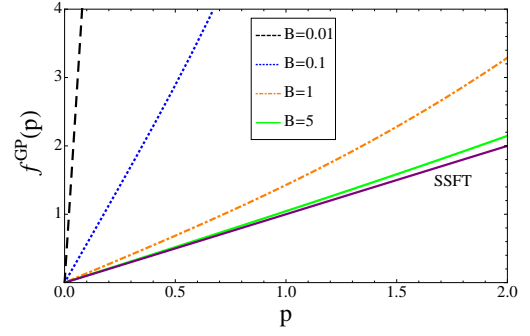} 
\caption{\label{Fig_ft_gp3}The asymptotic fluctuation function $f^{GP}(p)$, Eq.~(\ref{ff_gp}), plotted as a function of $p$ for various $B$ values in regime (iii). $f^{GP}_\tau(p)$ increases monotonically for large $p$. For small $B$ values the slope of $f^{GP}(p)$ becomes singular, while for large $B$ the SSFT is recovered (solid purple curve). Here, $p^*=0.5$.}
\end{center}
\end{figure}

(iii) $v<0$ with $\tau_p<\tau_\lambda$ (Fig.~\ref{Fig_ft_gp3}). In this parameter regime the fluctuation function $f(p)$ is not defined because there are no negative work fluctuations. However, under the influence of the Gaussian noise negative work fluctuations do arise, as discussed above (cf. Fig.~\ref{Fig_rf_gp3}) and one can discuss the properties of $f^{GP}(p)$. For large $B$, $f^{GP}(p)$ approaches the SSFT as in the other two cases. For small $B$, $f^{GP}(p)$ becomes steeper the smaller the $B$ value and in fact diverges as $B\rightarrow\infty$. This is consistent with a vanishing probability of observing negative work (cf. Eq.~\ref{gp_ff_def}), because the fluctuation function, defined by Eq.~(\ref{gp_ff_def}), diverges for $\Pi^{GP}_\tau(-p)\rightarrow 0$.

\section{Concluding remarks}
\label{Sec_C}

(1) As one of our main results we have shown that the work distribution in our two models --- the dragged particle with purely PSN and with PSN plus additional (thermal) Gaussian noise --- exhibits a large deviation form but that the SSFT does not hold for general values of the parameters. This result differs from previous studies of the nonequilibrium particle model considered here, where the noise statistics was taken as white Gaussian noise \cite{VanZon03} or white L\'evy noise \cite{Touchette07,Touchette09}, respectively. In the Gaussian case the work distribution has a large deviation form, which satisfies the SSFT, while in the L\'evy case there is no large deviation form and the SSFT does not hold. The PSN case is in this sense intermediate between the Gaussian and the L\'evy cases.

(2) One of the striking signatures of the work fluctuations in our two models is a parameter regime where the fluctuation function is strongly negative. This feature is due to the asymmetric tails of the work distribution for $v>0$: if the noise is purely PSN the left tail decays exponentially while the right tail decays faster than exponential to the positive work cut-off, where $\Pi_\tau(p^*)=0$. If we superimpose Gaussian noise the cut-off vanishes and the positive tail of the work distribution decays like a Gaussian. In that case the negative regime of the fluctuation function is even more pronounced and $f^{GP}(p)$ decreases monotonically for large $p$ (cf. Fig.~\ref{Fig_ft_gp1}). The unusually large negative fluctuations in our model could be useful for applications (cf. \cite{Ciliberto04}).

(3) We have considered the work fluctuations in the asymptotic regime of the NESS. Other well-known results of nonequilibrium statistical mechanics are the \textit{transient} fluctuation theorems (TFTs) \cite{EvansD94,Crooks99} and the nonequilibrium work relation (Jarzysnki relation, JR) \cite{Jarzynski97}, which could be checked in our model in the case of PSN plus thermal Gaussian noise. The TFTs and the JR only apply if the initial condition of the work measurement is sampled from a thermal equilibrium state (the JR requires in addition that the system equilibrates at the end of the work measurement).

It is then important to note that in our model the thermal equilibrium state is not achieved simply by setting $v=0$. In fact, the distribution of particle positions at $v=0$, $P_{GP}(y)$ of Eq.~(\ref{conv}), is evidently different from the Boltzmann equilibrium distribution. In order to generate an equilibrium initial state one would have to assume that initially the system is only coupled to a thermal heat bath and that the PSN is only `turned on' after the potential begins to move.

(4) Several variations of the PSN that we consider here are possible. For example one could assume a pulse shape different from the delta-peaks of Eq.~(\ref{F_poiss_noise}) or consider other forms of the distribution of amplitudes $\rho(\Gamma)$. In particular, one could consider symmetric PSN which acts \textit{double-sided}, e.g., by choosing a Gaussian distribution for $\rho(\Gamma)$ or by adding two PSN of the form of Eq.~(\ref{F_poiss_noise}), one with strictly positive amplitudes $\Gamma$ and one with strictly negative amplitudes, respectively. One would expect that those features of our model that are based on the asymmetry of the noise would disappear under such symmetric double-sided PSN. This will be discussed in more detail in a forthcoming publication.

(5) Our model with purely PSN represents an effectively nonlinear system due to the infinite barrier in the potential induced by the noise. A genuinely nonlinear system modulated by a periodic field has been investigated in \cite{Dykman08} and has also been shown to violate the SSFT.

(6) Our results might be relevant in the context of analytic frameworks for studying gene expression. Similar models as ours have been considered in \cite{Paulsson00,Friedman06,Azaele09} to model the stochastic dynamics of protein concentration in a cell. Here, the PSN represents random bursts in protein production, which occur with an exponentially distributed number of molecules.

(7) PSN occurs quite naturally in electric circuits, where the discreteness of the electron charge causes time-dependent fluctuations of the electric current. Our theory could thus be realized in an experiment similar to the resistance-capacitor dipole of \cite{VanZon04b,Garnier05}. If the Brownian Johnson-Nyquist noise is sufficiently weak compared with the shot noise one might be able to observe the strongly negative regime of the fluctuation function (cf. Fig.~\ref{Fig_ft_gp1}).

(8) We note that our theory could also be adapted to an experiment similar to that of Mahadevan \textit{et al} \cite{Mahadevan03}, where a lubricated rod of a hydrogel sliding on a soft vibrating substrate is considered as a model for biomimetic ratcheting motion. Instead of the purely oscillatory external vibrations of \cite{Mahadevan03} one could induce external asymmetric PSN, as considered here, which could then lead to work fluctuations with similar features as those presented here.

(9) Our dragged particle model under PSN and Gaussian noise could also be realized experimentally by a micron-sized colloidal particle in water confined by a laser trap \cite{Ciliberto}, similar to the setup in \cite{Wang02,Blickle06,GomezSolano09}. Here, the fluid environment is the origin of the thermal noise, as in the case of the standard Brownian motion. Additionally, stochastic pulses in the form of PSN can be imposed on the colloidal particle using modulations of the laser beam. In this setup the laser beam gives rise to both the confining moving potential and the shot noise. It would then be interesting to see if the negative fluctuation function predicted by our theory could indeed be measured.

\begin{acknowledgements}

The authors thank H. Touchette and S. Ciliberto for stimulating discussions. They also gratefully acknowledge financial support of the National Science Foundation under award PHY-0501315.

\end{acknowledgements}

\begin{appendix}

\section{The C\'aceres-Budini theorem}
\label{App_cb}

Consider the generalized multi-component Ornstein-Uhlenbeck process
\begin{eqnarray}
\label{app_eq}
\mathbf{\dot{x}}(t)=\mathcal{B}\mathbf{x}(t)+\boldsymbol{\zeta}(t),
\end{eqnarray}
where $\mathbf{x}(t)$ and $\boldsymbol{\zeta}(t)$ are $n$-component column vectors and $\mathcal{B}$ is an $n\times n$ square matrix. Let us assume that the characteristic functional of the noise $\boldsymbol{\zeta}(t)$ is known,
\begin{eqnarray}
\label{app_noise}
G_{\boldsymbol{\zeta(t)}}[\mathbf{k}(t)]\equiv\left<e^{i\int_0^\infty \mathbf{k}(t)\cdot\boldsymbol{\zeta}(t)\upd t}\right>,
\end{eqnarray}
where $\cdot$ denotes a scalar product and $\mathbf{k}(t)$ is defined as an $n$-component column vector (`test function'). The C\'aceres-Budini theorem then states that the characteristic functional of the process $\mathbf{x}(t)$ is obtained from $G_{\boldsymbol{\zeta(t)}}[\mathbf{k}(t)]$ according to
\begin{eqnarray}
\label{app_cb}
G_{\mathbf{x(t)}}[\mathbf{h}(t)]=e^{i\mathbf{k}_0\cdot\mathbf{x}_0}G_{\boldsymbol{\zeta}(t)}[\mathbf{k}(t)].
\end{eqnarray}
where $\mathbf{x}_0$ contains the initial conditions and the functional $\mathbf{k}(t)$ is related to $\mathbf{h}(t)$ via
\begin{eqnarray}
\label{app_kt}
\mathbf{k}(t)&=&\int_t^\infty e^{(s-t)\mathcal{B}^T}\mathbf{h}(s)\upd s,
\end{eqnarray}
and $\mathbf{k}_0\equiv\mathbf{k}(t=0)$. The process $\mathbf{x}(t)$ is therefore completely specified by $G_{\boldsymbol{\zeta}}[\mathbf{k}(t)]$, in particular, all cumulants of $\mathbf{x}(t)$ are obtained by functional derivation of the rhs of Eq.~(\ref{app_cb}) with respect to the components of $\mathbf{h}(t)$.

This theorem follows upon substitution of $\boldsymbol{\zeta}(t)=\mathbf{\dot{x}}(t)-\mathcal{B}\mathbf{x}$ from Eq.~(\ref{app_eq}) into Eq.~(\ref{app_noise}):
\begin{eqnarray}
G_{\boldsymbol{\zeta}}[\mathbf{k}(t)]=\left<e^{i\int_0^\infty \mathbf{k}(t)\cdot(\mathbf{\dot{x}}(t)-\mathcal{B}\mathbf{x}(t))\upd t}\right>.
\end{eqnarray}
Partial integration yields
\begin{eqnarray}
G_{\boldsymbol{\zeta}(t)}[\mathbf{k}(t)]&=&e^{-i\mathbf{k}_0\mathbf{x}_0}\left<e^{i\int_0^\infty(- \mathbf{\dot{k}}(t)-\mathcal{B}^T\mathbf{k}(t))\cdot\mathbf{x}(t)\upd t}\right>\nonumber\\
&=&e^{-i\mathbf{k}_0\mathbf{x}_0}\left<e^{i\int_0^\infty\mathbf{h}(t)\cdot\mathbf{x}(t)\upd t}\right>\nonumber\\
&=&e^{-i\mathbf{k}_0\mathbf{x}_0}G_{\boldsymbol{x}(t)}[\mathbf{h}(t)],
\end{eqnarray}
where in the second line we have set
\begin{eqnarray}
\label{app_h_k}
\mathbf{h}(t)=-\mathbf{\dot{k}}(t)-\mathcal{B}^T\mathbf{k}(t).
\end{eqnarray}
Consequently, $\mathbf{k}(t)$ is given as solution of Eq.~(\ref{app_h_k}), which reads
\begin{eqnarray}
\mathbf{k}(t)=e^{-t\mathcal{B}^T}\mathbf{k}_0-e^{-t\mathcal{B}^T}\int_0^te^{s\mathcal{B}^T}\mathbf{h}(s)\upd s.
\end{eqnarray}
Here, the initial condition $\mathbf{k}_0$ has to be chosen such that $\lim_{t\rightarrow\infty}\mathbf{k}(t)=\mathbf{0}$.

\section{NESS distributions of the particle position}

In this Appendix we determine the distribution of the particle position in the NESS for two cases: (i) The noise is given by PSN. (ii) The noise is given by a superposition of PSN and thermal Gaussian noise. We apply two different calculation methods. In (i) we solve the Fokker-Planck equation corresponding to the Langevin equation for $y(t)$. In (ii) we use the characteristic functional $G_{y(t)}[h(t)]$.

\subsection{NESS distribution for PSN}
\label{App_ss}

The Fokker-Planck equation for the distribution $p(y,t)$ corresponding to the Langevin equation~(\ref{F_y}) reads \cite{VanKampen}:
\begin{eqnarray}
\frac{\partial}{\partial t}p(y,t)&=&\frac{\partial}{\partial y}\left(\frac{1}{\tau_r}y+v_e\right)p(y,t)\nonumber\\
&&+\lambda \int_{-\infty}^\infty \rho(\Gamma)\left[p(y-\Gamma/\alpha,t)-p(y,t)\right]\upd \Gamma,\nonumber\\
\end{eqnarray}
where $\rho(\Gamma)$ is the distribution of the pulse amplitudes $\Gamma$. If we write the shift $-\Gamma/\alpha$ in the second term with the help of the shift operator $\exp\{-(\Gamma/\alpha)\partial/\partial y\}$ and perform the integral using the exponential distribution of amplitudes Eq.~(\ref{rho}), we obtain 
\begin{eqnarray}
\label{App_fp}
\frac{\partial}{\partial t}p(y,t)&=&\frac{\partial}{\partial y}\left(\frac{1}{\tau_r}y+v_e\right)p(y,t)\nonumber\\&&-\lambda\frac{\Gamma_0}{\alpha}\frac{\partial}{\partial y}\left(\frac{1}{1+\frac{\Gamma_0}{\alpha}\partial/\partial y}\right)p(y,t).
\end{eqnarray}
A solution of this differential equation can be found as follows. In the stationary state we have $\partial p(y,t)/ \partial t=0$. Furthermore, since $y(t)$ is a stationary process in a confining potential, the probability current vanishes as well. The Fokker-Planck equation~(\ref{App_fp}) therefore simplifies to
\begin{eqnarray}
\left(\frac{1}{\tau_r}y+v_e\right)P(y)-\lambda\Gamma_0\left(\frac{1}{1+\frac{\Gamma_0}{\alpha}\partial/\partial y}\right)P(y)=0.
\end{eqnarray}
We now define
\begin{eqnarray}
f(y)\equiv\left(\frac{1}{1+\frac{\Gamma_0}{\alpha}\partial/\partial y}\right)P(y),
\end{eqnarray}
i.e., $P(y)$ is obtained from $f(y)$ via
\begin{eqnarray}
\label{App_py}
P(y)=\left(1+\frac{\Gamma_0}{\alpha}\frac{\partial}{\partial y}\right)f(y).
\end{eqnarray}
In turn, the equation for $f(y)$ is only of first order:
\begin{eqnarray}
\left(\frac{1}{\tau_r}y+v_e\right)\left(1+\frac{\Gamma_0}{\alpha}\frac{\partial}{\partial y}\right)f(y)-\lambda\frac{\Gamma_0}{\alpha}f(y)=0.
\end{eqnarray}
The function $f(y)$ is thus given as
\begin{eqnarray}
f(y)\propto e^{-\int_{\bar{y}}^y\frac{y'+v\tau_r}{(y'+v_e\tau_r)\Gamma_0/\alpha}\upd y'},
\end{eqnarray}
where $\bar{y}$ denotes the lower integration limit. From Eq.~(\ref{App_fp} we then obtain a result for the distribution of $y$ in the NESS:
\begin{eqnarray}
P(y)&\propto&\frac{\lambda\Gamma_0\tau_r}{\alpha(y-y^*)}e^{-\int^y_{\bar{y}} \frac{\alpha(y'-y^*)-\lambda\Gamma_0\tau_r}{(y'-y^*)\Gamma_0}\upd y'},
\end{eqnarray}
where $y^*$ is the minimal value of the position in the steady state: $y^*=-v_e\tau_r$ (see Eq.~(\ref{min_pos})) and the lower integration limit $\bar{y}$ has to be chosen $\bar{y}>y^*$ for the integral to be well defined. Setting $\bar{y}=-v\tau_r$ then yields
\begin{eqnarray}
P(y)&\propto&\left(\frac{\alpha(y-y^*)}{\lambda\Gamma_0\tau_r}\right)^{\lambda\tau_r-1} e^{-y\alpha/\Gamma_0}.
\end{eqnarray}
The normalization constant can be calculated in a straightforward way and leads to the final result for the stationary distribution
\begin{eqnarray}
\label{p_stat_app}
P(y)=\frac{1}{\Gamma(\lambda\tau_r)}\frac{\alpha}{\Gamma_0}\left(\frac{\alpha}{\Gamma_0}(y-y^*)\right)^{\lambda\tau_r-1} e^{-(y-y^*)\alpha/\Gamma_0},\nonumber\\
\end{eqnarray}
where $\Gamma(x)$ denotes the Gamma function \cite{Abramowitz}. This result is used in Sec.~\ref{Sec_FT} in order to sample the initial condition in the steady state. Furthermore, we note that the Fourier transform of $P(y)$ is given by
\begin{eqnarray}
\label{Py_FT}
\mathcal{F}\left\{P(y)\right\}=\left(1-i\frac{\Gamma_0}{\alpha}\omega\right)^{-\lambda\tau_r}e^{i \omega y^*},
\end{eqnarray}
where $\omega$ is the Fourier variable conjugated to $y$.

\subsection{Steady state distribution for PSN with additional thermal Gaussian noise}
\label{App_gp}

The characteristic functional of $y(t)$ is defined as
\begin{eqnarray}
\label{y_cf_pg1}
G_{y(t)}[h(t)]\equiv \left<\exp\left\{i\int_0^\infty y(t)h(t)\upd t\right\}\right>.
\end{eqnarray}
From $G_{y(t)}[h(t)]$ the characteristic function of the particle position is obtained if we choose the test function
\begin{eqnarray}
\label{App_h1}
h_1(t)\equiv h_1\delta(t-t_1),
\end{eqnarray}
since substituting Eq.~(\ref{App_h1}) into Eq.~(\ref{y_cf_pg1}) yields
\begin{eqnarray}
G_{y(t)}[h_1(t)]=\left<e^{i h_1y(t_1)}\right>\equiv G_y(h_1,t_1),
\end{eqnarray}
where the rhs is just the definition of the characteristic function of the position $y$. Substituting $h_1(t)$ in the expression for $G_{y(t)}[h(t)]$, Eq.~(\ref{y_cf_pg2}), and taking the $t_1\rightarrow\infty$ limit yields
\begin{eqnarray}
G_y(h_1)&=&\left(1-i\frac{\Gamma_0}{\alpha}h_1\right)^{-\lambda\tau_r}e^{i h_1 y^*-\frac{\tau_r}{2\alpha\beta}h_1^2}.
\end{eqnarray}
The NESS distribution $P_{GP}(y)$ is the inverse Fourier transform of $G_y(h_1)$. Noting that $G_y(h_1)$ can be written as a product of the Fourier transform of $P(y)$, Eq.~(\ref{Py_FT}), and a Gaussian, we can express the distribution $P_{GP}(y)$ as the convolution integral
\begin{eqnarray}
\label{conv_app}
P_{GP}(y)&=& \int_{-\infty}^\infty P(y')P_G(y-y')\upd y',
\end{eqnarray}
where $P(y)$ is given by Eq.~(\ref{p_stat_app}), and $P_G(y)$ is given by the Gaussian
\begin{eqnarray}
P_G(y)=\sqrt{\frac{\beta\kappa}{2\pi}}e^{-\frac{\beta\kappa}{2}y^2}.
\end{eqnarray}
The convolution integral in Eq.~(\ref{conv_app}) can be evaluated in closed form \cite{Mathematica} and yields:
\begin{widetext}
\begin{eqnarray}
\label{pg_dist_closed}
P_{GP}(y)&=&\frac{1}{\Gamma(\lambda\tau_r)}\frac{\alpha}{\Gamma_0}\sqrt{\frac{\beta\kappa}{2\pi}}\frac{1}{2}\left(\frac{\beta\kappa \Gamma_0^2}{2\alpha^2}\right)^{-(\lambda\tau_r+1)/2}e^{-\frac{\beta\kappa}{2}(y-y^*)^2}\nonumber\\
&&\times\left[\sqrt{\frac{\beta\kappa\Gamma_0^2}{2\alpha^2}}\Gamma\left(\frac{\lambda\tau_r}{2}\right) {_1}F_1\left(\frac{\lambda\tau_r}{2},\frac{1}{2},\frac{1}{2}\left(\sqrt{\beta\kappa}(y-y^*)-\frac{\alpha}{\Gamma_0\sqrt{\beta\kappa}}\right)^2\right)\right.\nonumber\\
&&\left.+\left(\beta\kappa\frac{\Gamma_0}{\alpha}(y-y^*)-1\right)\Gamma\left(\frac{\lambda\tau_r+1}{2}\right){_1}F_1\left(\frac{\lambda\tau_r+1}{2},\frac{3}{2},\frac{1}{2}\left(\sqrt{\beta\kappa}(y-y^*)-\frac{\alpha}{\Gamma_0\sqrt{\beta\kappa}}\right)^2\right)\right],
\end{eqnarray}
\end{widetext}
where ${_1}F_1$ denotes the confluent hypergeometric function of the first kind \cite{Abramowitz} and $\Gamma(x)$ the Gamma function.

\section{Calculation of the work distribution for PSN and Gaussian noise}

\subsection{Saddle-point approximation and the rate function}
\label{App_sp}

The characteristic function of the work $G_{W_\tau}(q)$ is defined as
\begin{eqnarray}
G_{W_\tau}(q)\equiv\left<e^{iqW_\tau}\right>.
\end{eqnarray}
We can calculate $G_{W_\tau}(q)$ by substituting the particular test function $\tilde{h}(t)$ of Eq.~(\ref{tilde_h}) into the characteristic functional of $y(t)$, defined by Eq.~(\ref{y_cf_pg1})
\begin{eqnarray}
G_{y(t)}[\tilde{h}(t)]= \left<e^{-i q v\kappa \int_0^\tau y(t)\upd t}\right>=G_{W_\tau}(q).
\end{eqnarray}
Here, the last step follows due to definition of the work $W_\tau$, Eq.~(\ref{work_par}).

The work distribution $\Pi^{GP}_\tau(p)$, is then obtained as the inverse Fourier-transform of $G_{W_\tau}$, Eq.~(\ref{gp_cf}), i.e.
\begin{eqnarray}
\label{app_iF}
\Pi^{GP}_\tau(p)=\frac{\left<W_\tau\right>}{2\pi}\int_{-\infty}^\infty G_{W_\tau}(q)e^{-iqp\left<W_\tau\right>}\upd q.
\end{eqnarray}
After substitution of Eq.~(\ref{gp_cf}) in Eq.~(\ref{app_iF}) we see that $\Pi_\tau(p)$ of Eq.~(\ref{app_iF}) can be written in the form
\begin{eqnarray}
\label{app_FT_int}
\Pi^{GP}_\tau(p)=\frac{\alpha v^2\tau}{2\pi}\int_{-\infty}^\infty\chi(q)e^{\tau h(q)}\upd q,
\end{eqnarray}
where the functions $\chi(q)$ and $h(q)$ are given by
\begin{eqnarray}
\chi(q)&\equiv&\left(1+iq\Gamma_0v\right)^{\frac{\tau_r/\tau_\lambda}{1+iq \Gamma_0 v}}\nonumber\\
&&\times \exp\left\{-iqp^*\alpha v^2\tau_r +\frac{3}{2\beta}\alpha v^2\tau_r\right\},
\end{eqnarray}
and
\begin{eqnarray}
\label{app_sp_h}
h(q)&\equiv& iq\alpha v^2 (p^*-p)-q^2\frac{\alpha v^2}{\beta}\nonumber\\
&&+\frac{1}{\tau_\lambda}\left(\frac{1}{1+iq \Gamma_0 v }-1\right),
\end{eqnarray}
respectively. 
For large $\tau$ the integral in Eq.~(\ref{app_FT_int}) will be dominated by its saddle-point $\bar{q}$, which is determined by the condition $h'(\bar{q})=0$. Straightforward algebra then yields \begin{eqnarray}
\label{app_gp_q}
\bar{q}=\frac{i}{\Gamma_0 v}r_p,
\end{eqnarray}
where $r_p$ is determined by solving the cubic equation
\begin{eqnarray}
\label{app_cubic}
2B\, r_p(r_p-1)^2-\left(\frac{p^*-p}{p^*-1}\right)(r_p-1)^2+1=0.
\end{eqnarray}
This cubic equation is analyzed in more detail in Appendix~\ref{App_root}.

The saddle-point approximation of $\Pi^{GP}_\tau(p)$ is given by
\begin{eqnarray}
\label{app_sp_approx}
\Pi^{GP}_\tau(p)\cong\frac{\alpha v^2}{\sqrt{2\pi}}\sqrt{\frac{\tau}{|h''(\bar{q})|}}\chi(\bar{q})e^{i\theta+\tau h(\bar{q})},
\end{eqnarray}
where $\theta$ denotes the angle between the deformed integration path and the real axis. The result for $\Pi^{GP}_\tau(p)$ after substitution of the appropriate saddle-point $\bar{q}$, Eq.~(\ref{app_gp_q}), is presented in Eq.~(\ref{gp_work_dist}).

From Eq.~(\ref{app_sp_approx}) one can directly derive an expression for the rate function $I^{GP}(p)$. A comparison of Eq.~(\ref{app_sp_approx}) with the large deviation form Eq.~(\ref{gp_ldev}) yields
\begin{eqnarray}
\label{app_rf}
I^{GP}(p)=-h(\bar{q}),
\end{eqnarray}
where $h(\bar{q})$ of Eq.~(\ref{app_sp_h}) is more precisely given as $h(\bar{q})=h(\bar{q}(p),p)$, i.e., $h(\bar{q})$ depends on the dimensionless work value $p$ via the saddle-point $\bar{q}(p)$ and via $p$ directly. It then follows from Eq.~(\ref{app_rf}) that
\begin{eqnarray}
\frac{\upd}{\upd p}I^{GP}(p)&=&-\frac{\upd}{\upd p}h(\bar{q}(p),p)\nonumber\\
&=&-\frac{\partial}{\partial \bar{q}}h(\bar{q}(p),p)\bar{q}'(p)-\frac{\partial}{\partial p}h(\bar{q}(p),p).
\end{eqnarray}
The first term vanishes due to the property of the saddle-point $h(\bar{q}(p),p)=0$. The slope of the rate function is therefore given by
\begin{eqnarray}
\label{app_slope}
\frac{\upd}{\upd p}I^{GP}(p)&=&-\frac{\partial}{\partial p}h(\bar{q}(p),p)\nonumber\\
&=&i\bar{q}\alpha v^2,
\end{eqnarray}
by differentiation of Eq.~(\ref{app_sp_h}). The fact that the slope of the rate function is given by the saddle-point is a general result of the theory of large deviations (cf. \cite{Touchette08}).

\subsection{Analysis of the cubic roots}
\label{App_root}

The cubic equation (\ref{app_cubic}) can be written in the normal form
\begin{eqnarray}
\label{app_cubic2}
r^3-\left(2+\frac{\gamma}{2B}\right)r^2+\left(1+\frac{\gamma}{B}\right) r+\frac{1-\gamma}{2B}=0,
\end{eqnarray}
where
\begin{eqnarray}
\label{app_gamma}
\gamma\equiv\frac{p^*-p}{p^*-1}.
\end{eqnarray}
Although we could use Cardano's formula to investigate the behavior of the roots as functions of $B$ and $\gamma$, we obtain the same information using simple calculus. We define the function
\begin{eqnarray}
f(r)\equiv r^3-\left(2+\frac{\gamma}{2B}\right)r^2+\left(1+\frac{\gamma}{B}\right) r+\frac{1-\gamma}{2B}.
\end{eqnarray}
Setting the derivative of $f(r)$ to zero leads to simple expressions for the location of the extrema of $f(r)$, given by
\begin{eqnarray}
\label{fr_ex}
r_1=1,\qquad,\qquad r_2=\frac{1}{3}+\frac{\gamma}{3B}
\end{eqnarray}
Since the coefficient in the cubic term of $f(r)$ is $>0$, $f(r)$ is monotonically decreasing for $r\rightarrow-\infty$ and monotonically increasing for $r\rightarrow\infty$. This implies that, if $r_2>r_1$, $f(r)$ has a maximum at $r_1$ and a minimum at $r_2$. On the other hand, if $r_2<r_1$, $f(r)$ has a maximum at $r_2$ and a minimum at $r_1$. By substitution we find that $f(r_1)=1/(2B)$, i.e., $f(r_1)$ is positive for all values of $B$. Consequently, also $f(r_2)>1/(2B)$ if $r_2<r_1$. From this information about the extrema of $f(r)$ we can conclude that $f(r)$ must intersect with the $r$-axis either when approaching the maximum at $r_1=1$, if $r_2>r_1$, or when approaching the maximum at $r_2$, if $r_2<r_1$. Therefore, there exists always a real root $\bar{r}<1$ of the cubic equation (\ref{app_cubic}), which is the appropriate root for the saddle-point $\bar{q}$, Eq.~(\ref{gp_q}), because the integration path in Eq.~(\ref{app_FT_int}) can be deformed to go through $\bar{q}$ without crossing the pole in the characteristic function $G_{W_\tau}$, Eq.~(\ref{gp_cf}), at $q=i/(\Gamma_0 v)$.

Moreover, the behavior of this root under a change of $\gamma$ can be assessed qualitatively from this analysis. For large $\gamma$, $\bar{r}$ remains in the vicinity of $r_1$, so that $\bar{r}\rightarrow 1$ for $\gamma\rightarrow\infty$. On the other hand, for negative $\gamma$, $\bar{r}$ remains in the vicinity of $r_2$, which is proportional to $\gamma$ (cf. Eq.~(\ref{fr_ex})), so that qualitatively $\bar{r}\propto \gamma$ for $\gamma\rightarrow-\infty$. These results can also be obtained from Cardano's formula for the three roots of Eq.~(\ref{app_cubic}).

Since $\gamma$ is given by Eq.~(\ref{app_gamma}), where $p^*=1+\sigma(v)\tau_p/\tau_\lambda$, Eq.~(\ref{p_cutoff}), we thus find that

(i) $v>0$: $\bar{r}\rightarrow 1$ for $p\rightarrow-\infty$ and $\bar{r}\propto -p$ for $p\rightarrow\infty$.

(ii) $v<0$: $\bar{r}\rightarrow 1$ for $p\rightarrow\infty$ and $\bar{r}\propto p$ for $p\rightarrow-\infty$.

This behavior of the root $\bar{r}$ determines the properties of the tails of the rate function $I^{GP}(p)$ due to Eq.~(\ref{app_slope}).

\end{appendix}

%\bibliography{/Volumes/ADDIHD/MyFiles/Latex/Fluctuations.bib}

\begin{thebibliography}{42}
\expandafter\ifx\csname natexlab\endcsname\relax\def\natexlab#1{#1}\fi
\expandafter\ifx\csname bibnamefont\endcsname\relax
  \def\bibnamefont#1{#1}\fi
\expandafter\ifx\csname bibfnamefont\endcsname\relax
  \def\bibfnamefont#1{#1}\fi
\expandafter\ifx\csname citenamefont\endcsname\relax
  \def\citenamefont#1{#1}\fi
\expandafter\ifx\csname url\endcsname\relax
  \def\url#1{\texttt{#1}}\fi
\expandafter\ifx\csname urlprefix\endcsname\relax\def\urlprefix{URL }\fi
\providecommand{\bibinfo}[2]{#2}
\providecommand{\eprint}[2][]{\url{#2}}

\bibitem[{\citenamefont{Wang et~al.}(2002)\citenamefont{Wang, Sevick, Mittag,
  Searles, and Evans}}]{Wang02}
\bibinfo{author}{\bibfnamefont{G.~M.} \bibnamefont{Wang}},
  \bibinfo{author}{\bibfnamefont{E.~M.} \bibnamefont{Sevick}},
  \bibinfo{author}{\bibfnamefont{E.}~\bibnamefont{Mittag}},
  \bibinfo{author}{\bibfnamefont{D.~J.} \bibnamefont{Searles}},
  \bibnamefont{and} \bibinfo{author}{\bibfnamefont{D.~J.} \bibnamefont{Evans}},
  \bibinfo{journal}{Physical Review Letters} \textbf{\bibinfo{volume}{89}},
  \bibinfo{pages}{050601} (\bibinfo{year}{2002}).

\bibitem[{\citenamefont{{van Zon} and Cohen}(2003{\natexlab{a}})}]{VanZon03}
\bibinfo{author}{\bibfnamefont{R.}~\bibnamefont{{van Zon}}} \bibnamefont{and}
  \bibinfo{author}{\bibfnamefont{E.~G.~D.} \bibnamefont{Cohen}},
  \bibinfo{journal}{Physical Review E} \textbf{\bibinfo{volume}{67}},
  \bibinfo{pages}{046102} (\bibinfo{year}{2003}{\natexlab{a}}).

\bibitem[{\citenamefont{Taniguchi and Cohen}(2007)}]{Taniguchi07}
\bibinfo{author}{\bibfnamefont{T.}~\bibnamefont{Taniguchi}} \bibnamefont{and}
  \bibinfo{author}{\bibfnamefont{E.~G.~D.} \bibnamefont{Cohen}},
  \bibinfo{journal}{Journal of Statistical Physics}
  \textbf{\bibinfo{volume}{126}}, \bibinfo{pages}{1} (\bibinfo{year}{2007}).

\bibitem[{\citenamefont{Taniguchi and Cohen}(2008{\natexlab{a}})}]{Taniguchi08}
\bibinfo{author}{\bibfnamefont{T.}~\bibnamefont{Taniguchi}} \bibnamefont{and}
  \bibinfo{author}{\bibfnamefont{E.~G.~D.} \bibnamefont{Cohen}},
  \bibinfo{journal}{Journal of Statistical Physics}
  \textbf{\bibinfo{volume}{130}}, \bibinfo{pages}{1}
  (\bibinfo{year}{2008}{\natexlab{a}}).

\bibitem[{\citenamefont{Evans et~al.}(1993)\citenamefont{Evans, Cohen, and
  Morriss}}]{EvansD93}
\bibinfo{author}{\bibfnamefont{D.~J.} \bibnamefont{Evans}},
  \bibinfo{author}{\bibfnamefont{E.~G.~D.} \bibnamefont{Cohen}},
  \bibnamefont{and} \bibinfo{author}{\bibfnamefont{G.~P.}
  \bibnamefont{Morriss}}, \bibinfo{journal}{Physical Review Letters}
  \textbf{\bibinfo{volume}{71}}, \bibinfo{pages}{2401} (\bibinfo{year}{1993}).

\bibitem[{\citenamefont{Gallavotti and
  Cohen}(1995{\natexlab{a}})}]{Gallavotti95}
\bibinfo{author}{\bibfnamefont{G.}~\bibnamefont{Gallavotti}} \bibnamefont{and}
  \bibinfo{author}{\bibfnamefont{E.~G.~D.} \bibnamefont{Cohen}},
  \bibinfo{journal}{Physical Review Letters} \textbf{\bibinfo{volume}{74}},
  \bibinfo{pages}{2694} (\bibinfo{year}{1995}{\natexlab{a}}).

\bibitem[{\citenamefont{Gallavotti and
  Cohen}(1995{\natexlab{b}})}]{Gallavotti95b}
\bibinfo{author}{\bibfnamefont{G.}~\bibnamefont{Gallavotti}} \bibnamefont{and}
  \bibinfo{author}{\bibfnamefont{E.~G.~D.} \bibnamefont{Cohen}},
  \bibinfo{journal}{Journal of Statistical Physics}
  \textbf{\bibinfo{volume}{80}}, \bibinfo{pages}{931}
  (\bibinfo{year}{1995}{\natexlab{b}}).

\bibitem[{\citenamefont{Kurchan}(1998)}]{Kurchan98}
\bibinfo{author}{\bibfnamefont{J.}~\bibnamefont{Kurchan}},
  \bibinfo{journal}{Journal of Physics A: Mathematical and General}
  \textbf{\bibinfo{volume}{31}}, \bibinfo{pages}{3719} (\bibinfo{year}{1998}).

\bibitem[{\citenamefont{Lebowitz and Spohn}(1999)}]{Lebowitz99}
\bibinfo{author}{\bibfnamefont{J.~L.} \bibnamefont{Lebowitz}} \bibnamefont{and}
  \bibinfo{author}{\bibfnamefont{H.}~\bibnamefont{Spohn}},
  \bibinfo{journal}{Journal of Statistical Physics}
  \textbf{\bibinfo{volume}{95}}, \bibinfo{pages}{333} (\bibinfo{year}{1999}).
  
\bibitem{Footnote1}We refer to Eq.~(\ref{conventional}) as the steady state fluctuation theorem (SSFT), which, in fact, has the same form as the so-called Gallavotti-Cohen fluctuation theorem \cite{Gallavotti95,Gallavotti95b}. The SSFT has to be distinguished from similar theorems such as the Evans-Searles transient fluctuation theorem \cite{EvansD94} which is only valid for an initial equilibrium state. For a discussion of the relationship between the Gallavotti-Cohen and Evans-Searles fluctuation theorems, we refer to \cite{Cohen99}.

\bibitem[{\citenamefont{Evans and Searles}(1994)}]{EvansD94}
\bibinfo{author}{\bibfnamefont{D.~J.} \bibnamefont{Evans}} \bibnamefont{and}
  \bibinfo{author}{\bibfnamefont{D.~J.} \bibnamefont{Searles}},
  \bibinfo{journal}{Physical Review E} \textbf{\bibinfo{volume}{50}},
  \bibinfo{pages}{1645} (\bibinfo{year}{1994}).

\bibitem[{\citenamefont{Cohen and Gallavotti}(1999)}]{Cohen99}
\bibinfo{author}{\bibfnamefont{E.~G.~D.} \bibnamefont{Cohen}} \bibnamefont{and}
  \bibinfo{author}{\bibfnamefont{G.}~\bibnamefont{Gallavotti}},
  \bibinfo{journal}{Journal of Statistical Physics}
  \textbf{\bibinfo{volume}{96}}, \bibinfo{pages}{1343} (\bibinfo{year}{1999}).

\bibitem[{\citenamefont{Touchette and Cohen}(2007)}]{Touchette07}
\bibinfo{author}{\bibfnamefont{H.}~\bibnamefont{Touchette}} \bibnamefont{and}
  \bibinfo{author}{\bibfnamefont{E.~G.~D.} \bibnamefont{Cohen}},
  \bibinfo{journal}{Physical Review E} \textbf{\bibinfo{volume}{76}},
  \bibinfo{pages}{020101(R)} (\bibinfo{year}{2007}).

\bibitem[{\citenamefont{Touchette and Cohen}(2009)}]{Touchette09}
\bibinfo{author}{\bibfnamefont{H.}~\bibnamefont{Touchette}} \bibnamefont{and}
  \bibinfo{author}{\bibfnamefont{E.~G.~D.} \bibnamefont{Cohen}},
  \bibinfo{journal}{arXiv:0903.3869}  (\bibinfo{year}{2009}).

\bibitem[{\citenamefont{Feynman and Hibbs}(1965)}]{Feynman}
\bibinfo{author}{\bibfnamefont{R.~P.} \bibnamefont{Feynman}} \bibnamefont{and}
  \bibinfo{author}{\bibfnamefont{A.~R.} \bibnamefont{Hibbs}},
  \emph{\bibinfo{title}{{Quantum Mechanics and Path Integrals}}}
  (\bibinfo{publisher}{McGraw-Hill, New York}, \bibinfo{year}{1965}).

\bibitem[{\citenamefont{{Van Kampen}}(1992)}]{VanKampen}
\bibinfo{author}{\bibfnamefont{N.~G.} \bibnamefont{{Van Kampen}}},
  \emph{\bibinfo{title}{{Stochastic Processes in Physics and Chemistry}}}
  (\bibinfo{publisher}{North-Holland, Amsterdam}, \bibinfo{year}{1992}).

\bibitem[{\citenamefont{Baule and Cohen}(2009)}]{Baule09}
\bibinfo{author}{\bibfnamefont{A.}~\bibnamefont{Baule}} \bibnamefont{and}
  \bibinfo{author}{\bibfnamefont{E.~G.~D.} \bibnamefont{Cohen}},
  \bibinfo{journal}{Physical Review E} \textbf{\bibinfo{volume}{79}}, \bibinfo{eid}{030103(R)}
  (\bibinfo{year}{2009}).

\bibitem[{\citenamefont{Sekimoto}(1998)}]{Sekimoto98}
\bibinfo{author}{\bibfnamefont{K.}~\bibnamefont{Sekimoto}},
  \bibinfo{journal}{Progress of Theoretical Physics Supplement}
  \textbf{\bibinfo{volume}{130}}, \bibinfo{pages}{17} (\bibinfo{year}{1998}).

\bibitem[{\citenamefont{Taniguchi and
  Cohen}(2008{\natexlab{b}})}]{Taniguchi08b}
\bibinfo{author}{\bibfnamefont{T.}~\bibnamefont{Taniguchi}} \bibnamefont{and}
  \bibinfo{author}{\bibfnamefont{E.~G.~D.} \bibnamefont{Cohen}},
  \bibinfo{journal}{Journal of Statistical Physics}
  \textbf{\bibinfo{volume}{130}}, \bibinfo{pages}{633}
  (\bibinfo{year}{2008}{\natexlab{b}}).

\bibitem[{\citenamefont{Cohen}(2008)}]{Cohen08}
\bibinfo{author}{\bibfnamefont{E.~G.~D.} \bibnamefont{Cohen}},
  \bibinfo{journal}{Journal of Statistical Mechanics: Theory and Experiment}
  \textbf{\bibinfo{volume}{2008}}, \bibinfo{pages}{P07014}
  (\bibinfo{year}{2008}).

\bibitem[{\citenamefont{{van Kampen}}(1980)}]{VanKampen80}
\bibinfo{author}{\bibfnamefont{N.~G.} \bibnamefont{{van Kampen}}},
  \bibinfo{journal}{Physica A} \textbf{\bibinfo{volume}{102}},
  \bibinfo{pages}{489} (\bibinfo{year}{1980}).

\bibitem[{\citenamefont{Dykman et~al.}()\citenamefont{Dykman, Baule, and
  Cohen}}]{Dykman}
\bibinfo{author}{\bibfnamefont{M.~I.} \bibnamefont{Dykman}},
  \bibinfo{author}{\bibfnamefont{A.}~\bibnamefont{Baule}}, \bibnamefont{and}
  \bibinfo{author}{\bibfnamefont{E.~G.~D.} \bibnamefont{Cohen}},
  \bibinfo{journal}{unpublished} .

\bibitem[{\citenamefont{C{\'a}ceres and Budini}(1997)}]{Caceres97}
\bibinfo{author}{\bibfnamefont{M.~O.} \bibnamefont{C{\'a}ceres}}
  \bibnamefont{and} \bibinfo{author}{\bibfnamefont{A.~A.}
  \bibnamefont{Budini}}, \bibinfo{journal}{Journal of Physics A: Mathematical
  and General} \textbf{\bibinfo{volume}{30}}, \bibinfo{pages}{8427}
  (\bibinfo{year}{1997}).

\bibitem[{\citenamefont{Abramowitz and Stegun}(1972)}]{Abramowitz}
\bibinfo{author}{\bibfnamefont{M.}~\bibnamefont{Abramowitz}} \bibnamefont{and}
  \bibinfo{author}{\bibfnamefont{C.~A.} \bibnamefont{Stegun}},
  \emph{\bibinfo{title}{{Handbook of Mathematical Functions}}}
  (\bibinfo{publisher}{Dover, New York}, \bibinfo{year}{1972}).
  
\bibitem{Footnote3}Here and in the following figures we have chosen representative values for the parameters, which will determine the values of the position and work cut-offs.

\bibitem[{\citenamefont{Jeffreys and Jeffreys}(1956)}]{Jeffreys}
\bibinfo{author}{\bibfnamefont{H.}~\bibnamefont{Jeffreys}} \bibnamefont{and}
  \bibinfo{author}{\bibfnamefont{B.~S.} \bibnamefont{Jeffreys}},
  \emph{\bibinfo{title}{{Methods of Mathematical Physics}}}
  (\bibinfo{publisher}{Cambridge University Press, Cambridge},
  \bibinfo{year}{1956}).

\bibitem[{\citenamefont{{van Zon} and Cohen}(2004)}]{VanZon04}
\bibinfo{author}{\bibfnamefont{R.}~\bibnamefont{{van Zon}}} \bibnamefont{and}
  \bibinfo{author}{\bibfnamefont{E.~G.~D.} \bibnamefont{Cohen}},
  \bibinfo{journal}{Physical Review E} \textbf{\bibinfo{volume}{69}},
  \bibinfo{pages}{056121} (\bibinfo{year}{2004}).

\bibitem[{\citenamefont{Kim et~al.}(2007)\citenamefont{Kim, Lee, H{\"a}nggi,
  and Talkner}}]{Kim07}
\bibinfo{author}{\bibfnamefont{C.}~\bibnamefont{Kim}},
  \bibinfo{author}{\bibfnamefont{E.~K.} \bibnamefont{Lee}},
  \bibinfo{author}{\bibfnamefont{P.}~\bibnamefont{H{\"a}nggi}},
  \bibnamefont{and} \bibinfo{author}{\bibfnamefont{P.}~\bibnamefont{Talkner}},
  \bibinfo{journal}{Physical Review E} \textbf{\bibinfo{volume}{76}},
  \bibinfo{pages}{011109} (\bibinfo{year}{2007}).

\bibitem[{\citenamefont{Touchette}(2008)}]{Touchette08}
\bibinfo{author}{\bibfnamefont{H.}~\bibnamefont{Touchette}},
  \bibinfo{journal}{arXiv:0804.0327}  (\bibinfo{year}{2008}).

\bibitem{Footnote2}A restriction on the range of $p$-values is similar to that for deterministic Anosov systems, where the phase-space is bounded \cite{Gallavotti95,Gallavotti95b}.

\bibitem[{\citenamefont{Baule}(2008)}]{Baule}
\bibinfo{author}{\bibfnamefont{A.}~\bibnamefont{Baule}},
  \emph{\bibinfo{title}{{Exact results in driven stochastic systems}}}
  (\bibinfo{publisher}{University of Leeds PhD thesis}, \bibinfo{year}{2008}).
  
\bibitem{Footnote4}The calculations presented here for thermal Gaussian noise can be performed for a general Gaussian noise, with the only difference that then the noise strength is not related to the friction via a fluctuation-dissipation relation.

\bibitem[{\citenamefont{Ciliberto et~al.}(2004)\citenamefont{Ciliberto,
  Garnier, Hernandez, Lacpatia, Pinton, and {Ruiz Chavarria}}}]{Ciliberto04}
\bibinfo{author}{\bibfnamefont{S.}~\bibnamefont{Ciliberto}},
  \bibinfo{author}{\bibfnamefont{N.}~\bibnamefont{Garnier}},
  \bibinfo{author}{\bibfnamefont{S.}~\bibnamefont{Hernandez}},
  \bibinfo{author}{\bibfnamefont{C.}~\bibnamefont{Lacpatia}},
  \bibinfo{author}{\bibfnamefont{J.}~\bibnamefont{Pinton}}, \bibnamefont{and}
  \bibinfo{author}{\bibfnamefont{G.}~\bibnamefont{{Ruiz Chavarria}}},
  \bibinfo{journal}{Physica A} \bibinfo{pages}{240}
  (\bibinfo{year}{2004}).

\bibitem[{\citenamefont{Crooks}(1999)}]{Crooks99}
\bibinfo{author}{\bibfnamefont{G.~E.} \bibnamefont{Crooks}},
  \bibinfo{journal}{Physical Review E} \textbf{\bibinfo{volume}{60}},
  \bibinfo{pages}{2721} (\bibinfo{year}{1999}).

\bibitem[{\citenamefont{Jarzynski}(1997)}]{Jarzynski97}
\bibinfo{author}{\bibfnamefont{C.}~\bibnamefont{Jarzynski}},
  \bibinfo{journal}{Physical Review Letters} \textbf{\bibinfo{volume}{78}},
  \bibinfo{pages}{2690} (\bibinfo{year}{1997}).

\bibitem[{\citenamefont{Paulsson and Ehrenberg}(2000)}]{Paulsson00}
\bibinfo{author}{\bibfnamefont{J.}~\bibnamefont{Paulsson}} \bibnamefont{and}
  \bibinfo{author}{\bibfnamefont{M.}~\bibnamefont{Ehrenberg}},
  \bibinfo{journal}{Phys. Rev. Lett.} \textbf{\bibinfo{volume}{84}},
  \bibinfo{pages}{5447} (\bibinfo{year}{2000}).

\bibitem[{\citenamefont{Friedman et~al.}(2006)\citenamefont{Friedman, Cai, and
  Xie}}]{Friedman06}
\bibinfo{author}{\bibfnamefont{N.}~\bibnamefont{Friedman}},
  \bibinfo{author}{\bibfnamefont{L.}~\bibnamefont{Cai}}, \bibnamefont{and}
  \bibinfo{author}{\bibfnamefont{X.~S.} \bibnamefont{Xie}},
  \bibinfo{journal}{Physical Review Letters} \textbf{\bibinfo{volume}{97}},
  \bibinfo{eid}{168302} (\bibinfo{year}{2006}).

\bibitem[{\citenamefont{Azaele et~al.}(2009)\citenamefont{Azaele, Banavar, and
  Maritan}}]{Azaele09}
\bibinfo{author}{\bibfnamefont{S.}~\bibnamefont{Azaele}},
  \bibinfo{author}{\bibfnamefont{J.~R.} \bibnamefont{Banavar}},
  \bibnamefont{and} \bibinfo{author}{\bibfnamefont{A.}~\bibnamefont{Maritan}},
  \bibinfo{journal}{arXiv:0902.0941}  (\bibinfo{year}{2009}).

\bibitem[{\citenamefont{Dykman}(2008)}]{Dykman08}
\bibinfo{author}{\bibfnamefont{M.~I.} \bibnamefont{Dykman}},
  \bibinfo{journal}{Physical Review E} \textbf{\bibinfo{volume}{77}},
  \bibinfo{pages}{021123} (\bibinfo{year}{2008}).

\bibitem[{\citenamefont{{van Zon} et~al.}(2004)\citenamefont{{van Zon},
  Ciliberto, and Cohen}}]{VanZon04b}
\bibinfo{author}{\bibfnamefont{R.}~\bibnamefont{{van Zon}}},
  \bibinfo{author}{\bibfnamefont{S.}~\bibnamefont{Ciliberto}},
  \bibnamefont{and} \bibinfo{author}{\bibfnamefont{E.~G.~D.}
  \bibnamefont{Cohen}}, \bibinfo{journal}{Physical Review Letters}
  \textbf{\bibinfo{volume}{92}}, \bibinfo{pages}{130601}
  (\bibinfo{year}{2004}).

\bibitem[{\citenamefont{Garnier and Ciliberto}(2005)}]{Garnier05}
\bibinfo{author}{\bibfnamefont{N.}~\bibnamefont{Garnier}} \bibnamefont{and}
  \bibinfo{author}{\bibfnamefont{S.}~\bibnamefont{Ciliberto}},
  \bibinfo{journal}{Physical Review E} \textbf{\bibinfo{volume}{71}},
  \bibinfo{pages}{060101(R)} (\bibinfo{year}{2005}).

\bibitem[{\citenamefont{Mahadevan et~al.}(2003)\citenamefont{Mahadevan, Daniel,
  and Chaudhury}}]{Mahadevan03}
\bibinfo{author}{\bibfnamefont{L.}~\bibnamefont{Mahadevan}},
  \bibinfo{author}{\bibfnamefont{S.}~\bibnamefont{Daniel}}, \bibnamefont{and}
  \bibinfo{author}{\bibfnamefont{M.~K.} \bibnamefont{Chaudhury}},
  \bibinfo{journal}{Proceedings of the National Academy of Sciences}
  \textbf{\bibinfo{volume}{101}}, \bibinfo{pages}{23} (\bibinfo{year}{2003}).
  
\bibitem{Ciliberto}S. Ciliberto, private communication.

\bibitem[{\citenamefont{Blickle et~al.}(2006)\citenamefont{Blickle, Speck,
  Helden, Seifert, and Bechinger}}]{Blickle06}
\bibinfo{author}{\bibfnamefont{V.}~\bibnamefont{Blickle}},
  \bibinfo{author}{\bibfnamefont{T.}~\bibnamefont{Speck}},
  \bibinfo{author}{\bibfnamefont{L.}~\bibnamefont{Helden}},
  \bibinfo{author}{\bibfnamefont{U.}~\bibnamefont{Seifert}}, \bibnamefont{and}
  \bibinfo{author}{\bibfnamefont{C.}~\bibnamefont{Bechinger}},
  \bibinfo{journal}{Physical Review Letters} \textbf{\bibinfo{volume}{96}},
  \bibinfo{eid}{070603} (\bibinfo{year}{2006}).

\bibitem[{\citenamefont{{Gomez-Solano}
  et~al.}(2009)\citenamefont{{Gomez-Solano}, Petrosyan, Ciliberto, Chetrite,
  and Gawedzki}}]{GomezSolano09}
\bibinfo{author}{\bibfnamefont{J.~R.} \bibnamefont{{Gomez-Solano}}},
  \bibinfo{author}{\bibfnamefont{A.}~\bibnamefont{Petrosyan}},
  \bibinfo{author}{\bibfnamefont{S.}~\bibnamefont{Ciliberto}},
  \bibinfo{author}{\bibfnamefont{R.}~\bibnamefont{Chetrite}}, \bibnamefont{and}
  \bibinfo{author}{\bibfnamefont{K.}~\bibnamefont{Gawedzki}},
  \bibinfo{journal}{arXiv:0903.1075}  (\bibinfo{year}{2009}).

\bibitem[{Mat(2008)}]{Mathematica}
\emph{\bibinfo{title}{Mathematica}} (\bibinfo{publisher}{Wolfram Research
  Inc.}, \bibinfo{year}{2008}), \bibinfo{edition}{Version 7.0}.
  

\end{thebibliography}

\end{document}